\documentclass[10pt,letterpaper]{article}
\usepackage[top=0.85in,left=2.75in,footskip=0.75in]{geometry}

\usepackage{amsmath,amssymb}

\usepackage{changepage}

\usepackage[utf8x]{inputenc}

\usepackage{textcomp,marvosym}

\usepackage{cite}

\usepackage{nameref,hyperref}

\usepackage[right]{lineno}

\usepackage{microtype}
\DisableLigatures[f]{encoding = *, family = * }

\usepackage[table]{xcolor}

\usepackage{array}

\newcolumntype{+}{!{\vrule width 2pt}}

\newlength\savedwidth

\raggedright
\usepackage[section]{placeins}
\usepackage[aboveskip=1pt,labelfont=bf,labelsep=period,justification=raggedright,singlelinecheck=off]{caption}

\makeatletter
\renewcommand{\@biblabel}[1]{\quad#1.}
\makeatother

\date{}

\usepackage{lastpage,fancyhdr,graphicx}
\usepackage{epstopdf}
\fancyheadoffset[L]{2.25in}
\fancyfootoffset[L]{2.25in}
\newcommand\GR{{G_{\rm R}}}
\newcommand\GRprime{{\tilde{G}_{\rm R}}}
\newcommand\Pprime{{\tilde{P}}}
\newcommand\Grr{{G_{rr}}}
\newcommand\Gss{{G_{ss}}}
\newcommand\Grs{{G_{rs}}}
\newcommand\Gsr{{G_{sr}}}
\newcommand\Gqr{{G_{\rm qr}}}
\newcommand\Gpr{{G_{\rm pr}}}
\newcommand\Gqrd{{G_{\rm qrd}}}
\newcommand\Gqrnd{{G_{\rm qrnd}}}
\newcommand\Wqr{{W_{\rm qr}}}
\newcommand\Wpr{{W_{\rm pr}}}
\newcommand\Wrr{{W_{rr}}}

\begin{document}
\vspace*{0.2in}

\begin{flushleft}
{\Large
\textbf\newline{Capturing the influence of geopolitical ties from Wikipedia with reduced Google matrix} 
}
\newline
\\
Samer El Zant\textsuperscript{1},
Katia Jaffr\`es-Runser\textsuperscript{1}
Dima L Shepelyansky\textsuperscript{2},
\\
\bigskip
\textbf{1} Institut de Recherche en Informatique de Toulouse/INPT, Universit\'e de Toulouse, Toulouse, France
\\
\textbf{2} Laboratoire de Physique Th\'eorique du CNRS/IRSAMC, Universit\'e de Toulouse, Toulouse, France
\\
\bigskip

%
%





* samer.elzant@enseeiht.fr

\end{flushleft}
\section*{Abstract}
Interactions between countries originate from diverse aspects such as geographic proximity, trade, socio-cultural habits, language, religions, etc. Geopolitics studies the influence of a country's geographic space on its political power and its relationships with other countries. 
 This work reveals the potential of Wikipedia mining for geopolitical study. Actually, Wikipedia offers solid knowledge and strong correlations among countries by linking web pages together for different types of information (e.g. economical, historical, political, and many others).
The major finding of this paper is to show that meaningful results on the influence of country ties can be extracted from the hyperlinked structure of Wikipedia. 
We leverage a novel stochastic matrix representation of Markov chains of complex directed networks called the reduced Google matrix theory. For a selected small size set of nodes, the reduced Google matrix concentrates direct and indirect links of the million-node sized Wikipedia network into a small  Perron-Frobenius matrix keeping the PageRank probabilities of the global Wikipedia network. We perform a novel sensitivity analysis that leverages this reduced Google matrix to characterize the influence of relationships between countries from the global network. We apply this analysis to two chosen sets of countries (i.e. the set of 27 European Union countries and a set of 40 top worldwide countries).
We show that with our sensitivity analysis we can exhibit easily very meaningful information on geopolitics from five different Wikipedia editions (English, Arabic, Russian, French and German).   



\section*{Introduction}

Relationships between countries have always been of utmost interest to study for countries themselves as they have to be accounted for into any country's strategic and diplomatic plan. Studies are driven by observing the influence of a relationship between two countries on other countries from different perspectives listing economic exchanges, social changes, history, politics, religious, martial, regional as seen in ~\cite{eric}. 
The major finding of this paper is to show that meaningful results on geopolitics interactions could be extracted from Wikipedia for a given selection of countries. Therefore, it can be leveraged to provide a picture of countries relationships offering a new framework for geopolitical studies.
In ~\cite{sara}, Sara Javanmardi et al. show that even though anyone can edit a Wikipedia entry at any time,  the average article quality increases as it goes through various edits. Wikipedia's accuracy for its scientific entries has been proved by comparing it to Encyclopedia Britannica and to PDQ - NCI's Comprehensive Database in ~\cite{giles,malolan}. To sum up, Wikipedia has become the largest accurate reliable free online open source of knowledge.

Wikipedia is an interesting target domain for network analysts due to the hyperlinked structure that provides a direct relationship between web pages and topics.
Research on such networks has derived content-independent effective metrics to rank nodes and edges of the graph based on their relevance to a given criteria (clustering, importance ranking, etc.). In this study we concentrate on one of the most popular network analysis algorithms: the PageRank algorithm ~\cite{brin,langville}. PageRank algorithm could be seen as a Markov chain process relying on the definition of the so-called Google Matrix which describes the network interconnection.
For various language editions of Wikipedia it has been shown that the PageRank vector produces a reliable ranking of historical figures over 35 centuries of human history ~\cite{zhirov,eom1,eom2,eom3,ermann} and a solid Wikipedia ranking of world universities (WRWU) ~\cite{zhirov,lages}. It has been shown as well that the Wikipedia ranking of historical figures is in a good agreement with the well-known Hart ranking ~\cite{hart}, while the WRWU is in a good agreement with the Shanghai Academic ranking of world universities ~\cite{shanghai}.

This paper analyses the networks extracted from 5 language editions of Wikipedia to study the influence of countries on each other. 
 We proceed with this analysis for two sets of countries: $i)$ the 27 member states of the European Union and $ii)$ the top 40 countries according to English Wikipedia PageRank. For each Wikipedia language edition, we build a standard Wikipedia network representation as follows. 
Each webpage in Wikipedia is related to a clearly defined topic. On each page, there are hyperlinks pointing to other webpages of the same Wikipedia edition that are related to the topic of interest. As such, webpages are interconnected through directed links (i.e. hyperlinks), creating network of webpages. It is common to model this network as a directed graph where vertices represent all webpages and oriented edges represent the hyperlinks. This graph is complex as it can hold up to several millions of vertices and about ten times more edges. 
   
This Wikipedia network graph models the direct links between topics. However, indirect links exist as well as two topics can be related by intermediary webpages. For instance, the webpage of France is indirectly related to Latvia because it has a direct link to the \emph{Environmental Performance Index} webpage, that contains a link to Latvia.
To correctly determine the interaction between the countries of interest, both direct and indirect interactions must be accounted for in our study. 

A solution that captures the full contribution of direct and indirect interactions within a single stochastic matrix representation of the network of webpages has been proposed in \cite{frahm}. This solution relies on the Google Matrix representation \cite{brin,langville} of the Wikipedia network of $N$ nodes. Knowing a selection of $N_r$ nodes, $N_r \ll N$, it calculates a \emph{reduced Google matrix}. In this paper, the $N_r$ nodes are the Wikipedia webpages whose topics are the countries of interest\footnote{for instance we have \url{https://en.wikipedia.org/wiki/France} for France, \url{https://en.wikipedia.org/wiki/United_States} for US, etc.}. The reduced Google matrix $\GR$ is a $N_r$-by-$N_r$ Perron-Frobenius matrix where each element $\GR(i,j)$ represents the probability that node $i$ points onto node $j$ using direct and indirect links in the complete Wikipedia network.
Moreover, the reduced Google matrix theory offers a matrix decomposition of $\GR$ that can be leveraged to distinguish the contribution of direct and indirect interactions from the overall $\GR(i,j)$ probability. 
Up to a constant multiplier, the PageRank probabilities of $\GR$ are the same 
as for the Google matrix $G$ of the global Wikipedia network. 

Reduced Google matrix theory has been successfully leveraged in~\cite{politwiki}  and in~\cite{geop}. Results in~\cite{politwiki} highlight meaningful interactions between groups of political leaders from the Wikipedia networks. Most relevant to the study presented in this paper, the work in ~\cite{geop} shows that reduced Google matrix is a perfect candidate for analyzing the geopolitics interactions between countries selected worldwide for 5 different Wikipedia language editions for two reasons: 1) Indirect interactions components of $\GR$ capture reasonable and relevant information about hidden relationships between countries identified as hidden friends and followers 2) Part of the interactions are cross-cultural while others are clearly biased by the culture of the authors. This work has assessed the validity of the reduced Google matrix approach for the study of geopolitical interactions. It has extracted meaningful pieces of information from the intrinsic structure of the Wikipedia network by revealing the existence of indirect relationships between countries. 

The work of this paper goes one step further as it quantifies the influence of a relationship between two countries on the rest of the reduced network using $\GR$. Previous work has identified the strongest ties, but this one focuses on capturing the impact of a change in the strength of a relationship between two countries on the overall network interactions of selected countries via the global network. The impact on the overall network structure is measured by calculating the variation of importance of the nodes in the network.   
We show that this sensitivity analysis renders a reasonable and meaningful idea of the influence of a given bilateral tie on the whole network.  

More specifically, in this paper, we calculate $\GR$ for the two groups of 27 EU  and 40 world countries each. Thus, $\GR$ reflects in a 40-by-40 or 27-by-27 matrix the complete (direct and indirect) relationships between countries. To identify the relative influence of one relationship between two nations, we propose in this paper to compute a logarithmic derivative of the PageRank probabilities calculated from $\GR$ and $\GRprime$. PageRank probabilities are derived from $\GR$ as explained later. They represent the importance of a node in the network. 
$\GRprime$ is almost equal to $\GR$. It only differs by the values of one column. If the relationship going from nation $j$ to nation $i$ is of interest in the study, only the values of column $j$ are changed to relatively inflate the probability $\GRprime(i,j)$ of nation $j$ ending in nation $i$ compared to the other ones. This is done in practice by increasing $\GRprime(i,j)$ and then normalizing the column again to unity as it required by the definition of the Google matrix.  

From our sensitivity analysis on both sets of countries, we extract reasonable and really interesting geopolitical influences. Indeed, for instance in the set of 27 EU countries, our data shows clearly that the Nordic group of nations (Sweden, Denmark, Finland) have strong relationships together. If one of them increases its ties to another EU country alone, the remaining ones see their importance drop. The same observation is made for the group created by Austria, Hungary and Slovenia nations. These observations have been made by geopolitical specialists as well in \cite{nordic} and \cite{slovenia}, respectively. 
Another striking result is the impact of the exit of Great Britain from EU on the other European countries. Our data shows that Ireland will be the most affected country, which is inline with a study delivered recently by the London School of Economics \cite{brexit}. 
From our worldwide set of 40 countries, we show that strengthening the relationship between Russia and the United States of America would negatively impact the importance of Ukraine worldwide, which is identical to the interpretation represented by Francis Fukuyama in a recent article ~\cite{francis}. 

The paper is constructed as follows.
At first we introduce the reduced Google matrix theory, together with a primer on Google Matrix and PageRank calculations. The reduced Google matrix is illustrated for both sets of 27 EU and 40 nations. 
Next, the methodology for our link sensitivity analysis is presented. A detailed analysis for the two groups of countries is given in the Results section that focuses on the sensitivity analysis of important relationships in the group. Results are first given and discussed for the set of 27 EU countries and then for the set of 40 worldwide nations. Finally, conclusions are drawn in the last section.


\section*{Google Matrix analysis of Wikipedia networks}
\subsection*{Data Description}\label{sec:matrices}
Our study focuses on the networks representing 5 different Wikipedia editions\footnote{Data collected mid February 2013.}  from the set of 24 analyzed in \cite{eom3}: EnWiki, ArWiki, RuWiki, DeWiki and FrWiki that contain 4.212, 0.203 , 0.966, 1.533 and 1.353 millions of articles each.
The selected countries are the 27 EU countries as of February 2013\footnote{Croatia joined in July 2013} and the 40 countries selected from the EnWiki network as the top 40 countries 
of the PageRank probability for the complete network.

\begin{figure}[!ht]
\centering
\includegraphics[scale=0.5]{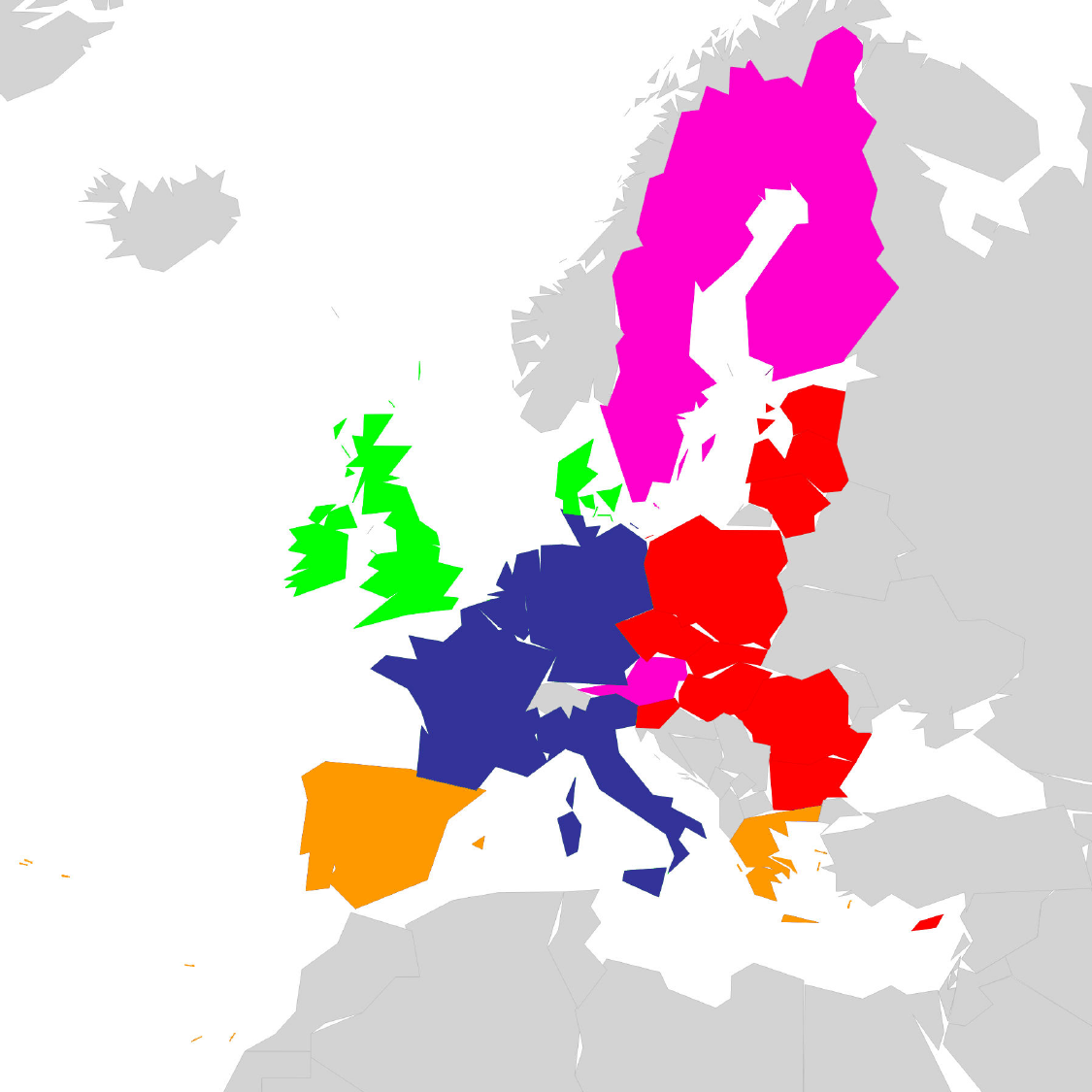}
\caption{{\bf Geographical distribution of the EU countries.}
Color code groups countries into 5 subsets: Blue (BL) for Founders, Green (GN) for 1973 new member states, Orange (OR) for 1981 to 1986 new member states, Pink (PK) for 1995 new member states and Red (RD) for 2004 to 2007 new member states.}
\label{Fig1}
\end{figure}

Countries that belong to the same region or having a common piece of history may probably exhibit stronger interactions in Wikipedia. For the set of 40 countries, we have created a color code that groups together countries that either belong to the same geographical region (e.g. Europe, South America, Middle East, North-East Asia, South-East Asia) or share a big part of history (former USSR; English speaking countries that are the legacy of the former British Empire)~\cite{geop}. On the other hand, EU countries are grouped upon their accession date to the union (e.g. Founder, 1973, 1981-1986, 1995, 2004-2007). Color code for EU countries can be seen in Fig~\ref{Fig1}. Color code for the worldwide set of 40 countries is available in Table~\ref{tab:countries}.

\begin{table}[!ht]
\centering
\caption{{\bf List of EU countries}}
\label{tab:EU}
\begin{tabular}{|c|c|c|c|c|c|c|c|c|}
\hline
\multicolumn{3}{|c|}{Wikipedia edition} & \multicolumn{2}{c|}{English} & \multicolumn{2}{c|}{French} & \multicolumn{2}{c|}{German} \\ \hline
Countries                	&	 CC	&	Color          	&	 K             	&	 K*           	&	 K            	&	 K*           	&	 K             	&	 K*           	\\	\hline
France                   	&	 FR   	&	 BL       	&	1	&	10	&	1	&	6	&	2	&	9	\\	\hline
United Kingdom           	&	 GB   	&	 GN      	&	2	&	14	&	4	&	13	&	24	&	27	\\	\hline
Germany                  	&	 DE   	&	 BL       	&	3	&	20	&	2	&	7	&	1	&	1	\\	\hline
Italy                    	&	 IT   	&	 BL       	&	4	&	6	&	3	&	9	&	4	&	14	\\	\hline
Spain                    	&	 ES   	&	 OR       	&	5	&	19	&	5	&	17	&	5	&	15	\\	\hline
Poland                   	&	 PL   	&	 RD        	&	6	&	3	&	8	&	5	&	6	&	6	\\	\hline
Netherlands              	&	 NL   	&	 BL       	&	7	&	25	&	7	&	12	&	7	&	21	\\	\hline
Sweden                   	&	 SE   	&	 PK       	&	8	&	13	&	11	&	25	&	8	&	18	\\	\hline
Romania                  	&	 RO   	&	 RD       	&	9	&	1	&	18	&	4	&	17	&	20	\\	\hline
Belgium                  	&	 BE   	&	 BL       	&	10	&	9	&	6	&	1	&	9	&	4	\\	\hline
Austria                  	&	 AT   	&	 PK       	&	11	&	27	&	9	&	23	&	3	&	3	\\	\hline
Greece                   	&	 GR   	&	 OR       	&	12	&	11	&	13	&	10	&	14	&	8	\\	\hline
Portugal                 	&	 PT   	&	 OR       	&	13	&	24	&	12	&	2	&	11	&	2	\\	\hline
Ireland                  	&	 IE   	&	 GN       	&	14	&	16	&	19	&	14	&	16	&	26	\\	\hline
Denmark                  	&	 DK   	&	 GN       	&	15	&	7	&	14	&	20	&	10	&	10	\\	\hline
Finland                  	&	 FI   	&	 PK       	&	16	&	4	&	17	&	18	&	15	&	7	\\	\hline
Hungary                  	&	 HU   	&	 RD       	&	17	&	2	&	10	&	3	&	13	&	12	\\	\hline
Czech Republic           	&	 CZ   	&	 RD       	&	18	&	5	&	15	&	24	&	12	&	17	\\	\hline
Bulgaria                 	&	 BG   	&	 RD       	&	19	&	22	&	20	&	11	&	20	&	13	\\	\hline
Estonia                  	&	 EE   	&	 RD       	&	20	&	8	&	24	&	15	&	22	&	23	\\	\hline
Slovenia                 	&	 SI   	&	 RD       	&	21	&	18	&	23	&	21	&	23	&	22	\\	\hline
Slovakia                 	&	 SK   	&	 RD       	&	22	&	12	&	16	&	8	&	18	&	5	\\	\hline
Lithuania                	&	 LT   	&	 RD       	&	23	&	21	&	22	&	27	&	21	&	19	\\	\hline
Cyprus                   	&	 CY   	&	 RD       	&	24	&	17	&	27	&	26	&	27	&	25	\\	\hline
Latvia                   	&	 LV   	&	 RD       	&	25	&	23	&	25	&	22	&	25	&	24	\\	\hline
Luxembourg               	&	 LU   	&	 BL       	&	26	&	26	&	21	&	19	&	19	&	11	\\	\hline
Malta                    	&	 MT   	&	 RD       	&	27	&	15	&	26	&	16	&	26	&	16	\\	\hline
\end{tabular}
\begin{flushleft}PageRank $K$ and CheiRank $K^*$ for EnWiki, FrWiki and DeWiki. Fig~\ref{Fig1} gives color correspondence details. Color code groups countries into 5 subsets: Blue (BL) for Founders, Green (GN) for 1973 new member states, Orange (OR) for 1981 to 1986 new member states, Pink (PK) for 1995 new member states and Red (RD) for 2004 to 2007 new member states. Standard country codes (CC) are given as well.
\end{flushleft}
\end{table}

\begin{table}[!ht]
\centering
\caption{{\bf List of 40 selected countries.}}
\label{tab:countries}
\begin{tabular}{|c|c|c|c|c|c|c|c|c|}
\hline
\multicolumn{3}{|c|}{Wikipedia edition} & \multicolumn{2}{c|}{English} & \multicolumn{2}{c|}{Arabic} & \multicolumn{2}{c|}{Russian} \\ \hline
Countries                & CC	&Color          & $K$          & $K^*$         & $K$         & $K^*$         & $K$          & $K^*$         \\ \hline
United States            & US   & OR        & 1            & 9             & 1           & 5             & 2            & 27            \\ \hline
France                   & FR   & RD       & 2            & 19            & 3           & 31            & 3            & 14            \\ \hline
United Kingdom           & GB   & OR       & 3            & 25            & 6           & 20            & 7            & 3             \\ \hline
Germany                  & DE   & RD       & 4            & 33            & 8           & 14            & 4            & 24            \\ \hline
Canada                   & CA   & OR       & 5            & 26            & 13          & 19            & 12           & 26            \\ \hline
India                    & IN   & PK       & 6            & 23            & 9           & 25            & 13           & 8             \\ \hline
Australia                & AU   & OR       & 7            & 35            & 16          & 22            & 18           & 12            \\ \hline
Italy                    & IT   & RD       & 8            & 15            & 5           & 1             & 6            & 32            \\ \hline
Japan                    & JP   & VT       & 9            & 4             & 11          & 9             & 11           & 7             \\ \hline
China                    & CN   & VT       & 10           & 8             & 12          & 17            & 9            & 21            \\ \hline
Russia                   & RU   & BL       & 11           & 6             & 7           & 2             & 1            & 2             \\ \hline
Spain                    & ES   & RD       & 12           & 30            & 4           & 8             & 8            & 15            \\ \hline
Poland                   & PL   & RD       & 13           & 12            & 26          & 32            & 10           & 17            \\ \hline
Netherlands              & NL   & RD       & 14           & 37            & 18          & 33            & 15           & 31            \\ \hline
Iran                     & IR   & YL       & 15           & 2             & 14          & 15            & 30           & 22            \\ \hline
Brazil                   & BR   & GN       & 16           & 3             & 21          & 26            & 20           & 1             \\ \hline
Sweden                   & SE   & RD       & 17           & 22            & 22          & 7             & 19           & 5             \\ \hline
New Zealand              & NZ   & OR       & 18           & 28            & 34          & 24            & 36           & 4             \\ \hline
Mexico                   & MX   & GN       & 19           & 40            & 23          & 38            & 22           & 37            \\ \hline
Switzerland              & CH   & RD       & 20           & 38            & 20          & 34            & 16           & 18            \\ \hline
Norway                   & NO   & RD       & 21           & 32            & 35          & 16            & 27           & 11            \\ \hline
Romania                  & RO   & RD       & 22           & 10            & 19          & 6             & 32           & 36            \\ \hline
Turkey                   & TR   & YL       & 23           & 7             & 15          & 13            & 21           & 38            \\ \hline
South Africa             & ZA   & OR       & 24           & 24            & 29          & 39            & 35           & 20            \\ \hline
Belgium                  & BE   & RD       & 25           & 18            & 27          & 37            & 29           & 30            \\ \hline
Austria                  & AT   & RD       & 26           & 39            & 28          & 28            & 14           & 28            \\ \hline
Greece                   & GR   & RD       & 27           & 21            & 10          & 36            & 25           & 25            \\ \hline
Argentina                & AR   & GN       & 28           & 1             & 32          & 29            & 33           & 23            \\ \hline
Philippines              & PH   & PK       & 29           & 17            & 36          & 21            & 39           & 33            \\ \hline
Portugal                 & PT   & RD       & 30           & 36            & 24          & 12            & 17           & 9             \\ \hline
Pakistan                 & PK   & PK       & 31           & 5             & 25          & 35            & 37           & 29            \\ \hline
Denmark                  & DK   & RD       & 32           & 16            & 33          & 10            & 31           & 19            \\ \hline
Israel                   & IL   & YL       & 33           & 20            & 17          & 18            & 28           & 6             \\ \hline
Finland                  & FI   & RD       & 34           & 14            & 37          & 4             & 26           & 16            \\ \hline
Egypt                    & EG   & YL       & 35           & 31            & 2           & 3             & 24           & 39            \\ \hline
Indonesia                & ID   & PK       & 36           & 13            & 31          & 11            & 34           & 10            \\ \hline
Hungary                  & HU   & RD       & 37           & 11            & 40          & 40            & 23           & 40            \\ \hline
Taiwan                   & TW   & VT       & 38           & 27            & 39          & 27            & 40           & 34            \\ \hline
South Korea              & KR   & VT       & 39           & 34            & 38          & 30            & 38           & 35            \\ \hline
Ukraine                  & UA   & BL       & 40           & 29            & 30          & 23            & 5            & 13            \\ \hline
\end{tabular}
\begin{flushleft}PageRank $K$ and CheiRank $K^*$ for EnWiki, FrWiki and RuWiki. Color code (CC) groups countries into 7 subsets: orange (OR) for English speaking countries, Blue (BL) for former Soviet union ones, Red (RD) for European ones, Green (GN) for South American ones, Yellow (YL) for Middle Eastern ones, Purple (VT) for North-East Asian ones and finally Pink (PK) for South-Eastern countries.
\end{flushleft}
\end{table}

\subsection*{Google matrix, PageRank and CheiRank.}

It is convenient to describe the network of $N$ Wikipedia articles by the Google matrix $G$ constructed from the adjacency matrix $A_{ij}$ with elements $1$ if article (node) $j$ points to  article (node) $i$ and zero otherwise. In this case, elements of the Google matrix take the standard form \cite{brin,langville}
\begin{equation}
  G_{ij} = \alpha S_{ij} + (1-\alpha) / N \;\; ,
\label{eq_gmatrix} 
\end{equation}
where $S$ is the matrix of Markov transitions with elements  $S_{ij}=A_{ij}/k_{out}(j)$, 
$k_{out}(j)=\sum_{i=1}^{N}A_{ij}\neq0$ being the node $j$ out-degree
(number of outgoing links) and with $S_{ij}=1/N$ if $j$ has no outgoing links (dangling node). 
Here $0< \alpha <1$ is the damping factor  which for a random surfer
determines the probability $(1-\alpha)$ to jump to any node;
below we use $\alpha=0.85$.
Element $G_{ij}$ represents the probability of a random surfer to go from node $j$ to node $i$.
The right eigenvectors $\psi_i(j)$  of $G$ are defined by:
\begin{equation}
\label{eq_gmatrix2}
\sum_{j'} G_{jj'} \psi_i(j')=\lambda_i \psi_i(j) \; .
\end{equation}

The PageRank eigenvector $P(j)=\psi_{i=0}(j) $ corresponds to the largest 
eigenvalue $\lambda_{i=0}=1$ \cite{brin,langville}. It has positive elements
which give the probability to find a random surfer on a given node
in the stationary long time limit of the Markov process.
All nodes can be ordered by a monotonically decreasing probability
$P(K_i)$ with the highest probability at $K=1$. The index $K$ 
is the PageRank index.  
Left eigenvectors are biorthogonal to right eigenvectors of different eigenvalues. 
The left eigenvector for $\lambda=1$ has identical (unit) entries  due to the column sum normalization of $G$. 
In the following we use the notations 
$\psi_L^T$  and $\psi_R$ for left and right eigenvectors, respectively. Notation $T$ stands for vector or matrix transposition.

In addition to the matrix $G$ it is useful to introduce a Google matrix $G^*$ constructed from the adjacency matrix of the same network but with inverted direction of all links \cite{linux}. 
The vector $P^*(K^*)$  is called the CheiRank vector \cite{linux,zhirov}, with $K^*$ being the CheiRank index obtained after numbering nodes in monotonic decrease of probability $P^*$. 
Thus, nodes with many ingoing (or outgoing) links have small values of $K=1,2,3...$ (or of $K^*=1,2,3,...$) \cite{langville,ermann}.

The two sets of 27 EU and 40 world countries are listed in Tables~\ref{tab:EU} and~\ref{tab:countries} respectively. The set of 40 countries has been chosen by selecting the countries with the largest PageRank probabilities in the full EnWiki network. 
In Table~\ref{tab:EU} and~\ref{tab:countries}, a \emph{local PageRank index} $K$ is given whose values range between $1$ and $27$ for EU countries, and between $1$ and $40$ for the other set. This local ranking keeps the countries in the same sequence as the original ranking over the entire network of webpages.  
The most influential countries are the top ranked ones with $K=1,2,...$. 
Similarly, the local CheiRank index $K^*$ \cite{linux,ermann} is given in both Tables for the two sets. At the top of $K^*$ we have the most communicative countries. Both local $K$ and $K^*$ are given for EnWiki, ArWiki and RuWiki. Not surprisingly, the order of top countries changes with respect to the edition (for instance, the top country for $K$ is US except for RuWiki whose top country is Russia).

\begin{figure}[!h]
\centering
\includegraphics[scale=0.7]{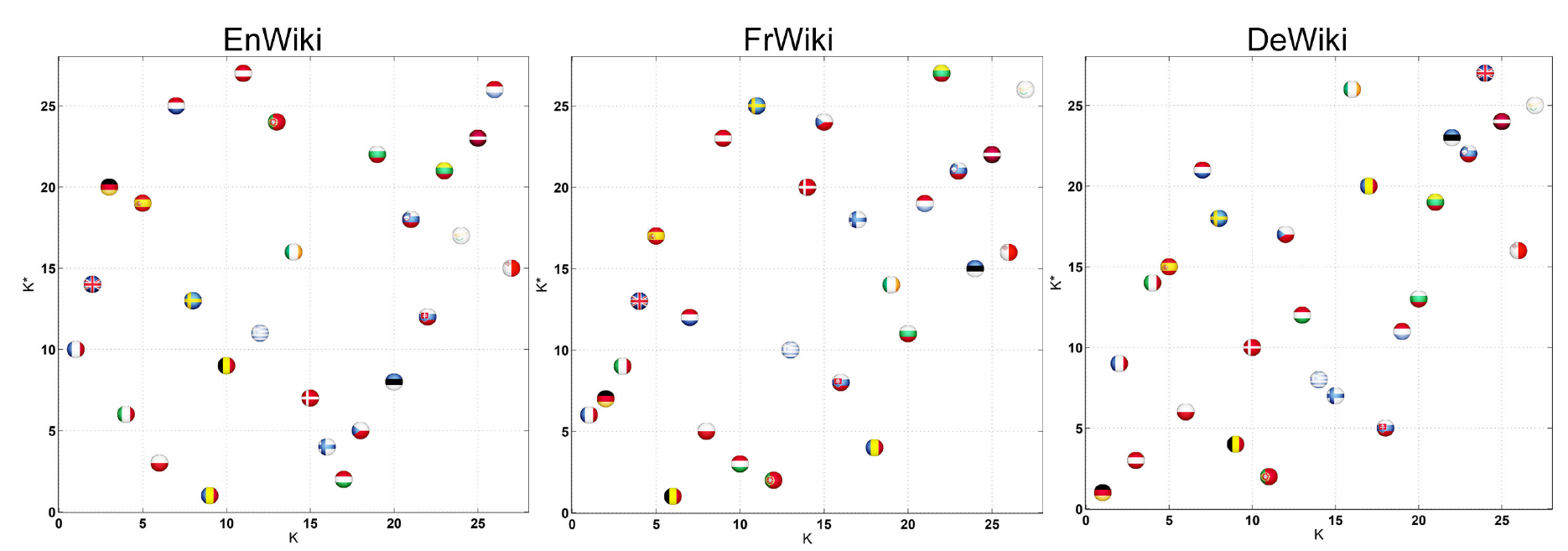}
\caption{{\bf Position of EU countries in the local $(K,K^*)$ plane.}
EnWiki (left), FrWiki (middle) and DeWiki (right) networks. Countries are marked by their flags.}
\label{Fig2}
\end{figure}

It is convenient as well to plot all nodes in the ($K$, $K^*$) plane to highlight the countries that are the most influential ($K=1, 2,$ ...) and the most communicative ($K^*=1, 2,$ ...) at the same time. Fig~\ref{Fig2} plots EU countries in the ($K$, $K^*$) plane for EnWiki, FrWiki and DeWiki editions. This plot is a bi-objective plot where $K$ and $K^*$ are to be minimized concurrently. It is interesting to look at the set of non-dominated countries which are the ones such that there is no other country beating them for both $K$ and $K^*$.

To summarize, PageRank and CheiRank capture the relative importance of nodes in the network. They are extracted from the Google matrix representation of the network of webpages. The Google matrix lists for each link the probability for directly transitioning from one webpage to the other one. The PagerRank probability $P(K)$ represents the probability of ending on a webpage, eventually. In the following, we introduce the Reduced Google Matrix that offers a complementary analysis that extracts the importance of the indirect interactions between a set of nodes of the original network.


\section*{Reduced Google Matrix analysis}
Let $G$ be the typical Google matrix of Eq~(\ref{eq_gmatrix}) for 
a network of $N$ nodes such that $G_{ij}\ge 0$ and the 
column sum normalization $\sum_{i=1}^N G_{ij}=1$ is verified for each column. 
We consider a sub-network 
with $N_r<N$ nodes, called ``reduced network''. In this case we can write 
$G$ in a block form:
\begin{equation}
\label{eq_Gblock}
G=\left(\begin{array}{cc}
\Grr & \Grs \\
\Gsr & \Gss \\
\end{array}\right)
\end{equation}
where the index ``$r$'' refers to the nodes of the reduced network and 
``$s$'' to the other $N_s=N-N_r$ nodes which form a complementary 
network which we will call the ``scattering network''. 
PageRank vector of the full network is given by:
\begin{equation}
\label{eq_Pageank0}
P=\left(\begin{array}{c}
P_r  \\
P_s  \\
\end{array}\right)
\end{equation}
which satisfies the equation $G\,P=P$ 
or in other words $P$ is the right eigenvector of $G$ for the 
unit eigenvalue. This eigenvalue equation reads in block notations:
\begin{eqnarray}
\label{eq_Pagerank1}
({\bf 1}-\Grr)\,P_r-\Grs\,P_s&=&0,\\
\label{eq_Pagerank2}
-\Gsr \,P_r+({\bf 1}-\Gss)\,P_s&=&0.
\end{eqnarray}
Here ${\bf 1}$ is the unit matrix of corresponding size $N_r$
or $N_s$.
Assuming that the matrix ${\bf 1}-\Gss$ is not singular, i.e. all 
eigenvalues $\Gss$ are strictly smaller than unity (in modulus), we obtain 
from Eq~(\ref{eq_Pagerank2}) that 
\begin{equation}
\label{eq_Ps}
P_s=({\bf 1}-\Gss)^{-1} \Gsr\,P_r
\end{equation}
which gives together with (\ref{eq_Pagerank1}):
\begin{equation}
\label{eq_Geff1}
\GR P_r=P_r\quad,\quad
\GR=\Grr+\Grs({\bf 1}-\Gss)^{-1} \Gsr
\end{equation}
where the matrix $\GR$ of size $N_r\times N_r$, defined for the 
reduced network, can be viewed as an effective reduced Google matrix. 
Here the contribution of $\Grr$ accounts for direct links 
in the reduced network and the second matrix inverse term 
corresponds to all contributions of indirect links of arbitrary order. 
The matrix elements of $\GR$ are non-negative since the matrix 
inverse in Eq~(\ref{eq_Geff1}) can be expanded as:
\begin{equation}
\label{eq_inverse_expand}
({\bf 1}-\Gss)^{-1}=\sum_{l=0}^\infty G_{ss}^{\,l} \;\; .
\end{equation}
In Eq~(\ref{eq_inverse_expand}) 
the integer $l$ represents the order of indirect links, i.~e. the number 
of indirect links which are used to connect indirectly two nodes of the 
reduced network. 
We refer the reader to \cite{politwiki} to get the proof that $\GR$ also fulfills the condition 
of column sum normalization being unity.

\subsection*{Numerical evaluation of $\GR$}
We can question how to evaluate practically the expression of Eq~(\ref{eq_Geff1}) of $\GR$ for a particular sparse and quite large 
network when $N_r\sim 10^2$-$10^3$ is small compared to $N$ and $N_s \approx N\gg N_r$. 
If $N_s$ is too large (e.~g. $N_s > 10^5$) a direct naive evaluation 
of the matrix inverse $({\bf 1}- \Gss)^{-1}$ in Eq~(\ref{eq_Geff1}) 
by Gauss algorithm is not efficient. In this case we can try the 
expansion of Eq~(\ref{eq_inverse_expand}) provided it converges sufficiently 
fast with a modest number of terms. However, this is most likely not the 
case for typical applications since $\Gss$ is very likely to have 
at least one eigenvalue very close to unity. 

Therefore, we consider the situation 
where the full Google matrix has a well defined gap between the leading 
unit eigenvalue and the second largest eigenvalue (in modulus). For example 
if $G$ is defined using a damping factor $\alpha$ in the standard way, 
as in Eq~(\ref{eq_gmatrix}), the 
gap is at least $1-\alpha$ which is $0.15$ for the standard choice 
$\alpha=0.85$ \cite{langville}.
In order to evaluate the expansion of Eq~(\ref{eq_inverse_expand}) efficiently, we need to take out analytically 
the contribution of the leading eigenvalue of $\Gss$ close to unity which is 
responsible for the slow convergence.

Below we denote by $\lambda_c$ this leading eigenvalue of $\Gss$ and by $\psi_R$ 
($\psi_L^T$) the corresponding right (left) eigenvector such that 
$\Gss\psi_R=\lambda_c\psi_R$ (or $\psi_L^T \Gss=\lambda_c\psi_L^T$). 
Both left and right eigenvectors as well as $\lambda_c$ can be efficiently 
computed by the power iteration method in a similar way as the standard 
PageRank method. 
Vectors $\psi_R$ are normalized with $E_s^T\psi_R=1$ and $\psi_L$ with $\psi_L^T\psi_R=1$. 
It is well known (and easy to show) that $\psi_L^T$ is orthogonal to all other 
right eigenvectors (and $\psi_R$ is orthogonal to all other 
left eigenvectors) of $\Gss$ with eigenvalues different from $\lambda_c$. 
We introduce the operator ${\cal P}_c=\psi_R\psi_L^T$ which is the 
projector onto the eigenspace of $\lambda_c$ and we denote by 
${\cal Q}_c={\bf 1}-{\cal P}_c$ the complementary projector. 
One verifies directly that both projectors commute with the matrix $\Gss$ 
and in particular ${\cal P}_c \Gss=\Gss{\cal P}_c=\lambda_c{\cal P}_c$. 
Therefore we can derive:
\begin{eqnarray}
\label{eq_inverse_project1}
({\bf 1}-\Gss)^{-1}&=
&{\cal P}_c\frac{1}{1-\lambda_c}+
{\cal Q}_c \sum_{l=0}^\infty \bar G_{ss}^{\,l}
\end{eqnarray}
with $\bar G_{ss}={\cal Q}_c \Gss{\cal Q}_c$ and using the 
standard identity ${\cal P}_c{\cal Q}_c=0$ for complementary 
projectors. 
The expansion in Eq~(\ref{eq_inverse_project1}) converges rapidly since $\bar G_{ss}^{\,l}\sim |\lambda_{c,2}|^l$ 
 with $\lambda_{c,2}$ being the second largest eigenvalue which is significantly lower than unity.

The combination of Eq~(\ref{eq_Geff1}) and Eq~(\ref{eq_inverse_project1}) 
provides an explicit algorithm feasible for a numerical implementation 
for modest values of $N_r$, large values of $N_s$ and of course 
if sparse matrices $G$, $\Gss$ are considered.  
We refer the reader to \cite{politwiki} for more advanced implementation considerations.

\subsection*{Decomposition of $\GR$}

On the basis of equations 
(\ref{eq_Geff1})-(\ref{eq_inverse_project1}),
the reduced Google matrix can be presented as a sum of three components:
\begin{equation}
\label{eq_3terms}
\GR=\Grr + \Gpr + \Gqr ,
\end{equation}
with the first component $\Grr$ given by direct matrix elements of $G$
among the selected $N_r$ nodes.
The second projector component $\Gpr$ is given by:
\begin{equation}
\label{eq_2ndterm}
\Gpr =  \Grs  {\cal P}_c \Gsr/(1-\lambda_c) , \; 
{\cal P}_c=\psi_R\psi_L^T \;.
\end{equation}

The third component $\Gqr$ is of particular interest in this study as it characterizes the impact of indirect or hidden links. It is given by:
\begin{equation}
\label{eq_3rdterm}
\Gqr =  \Grs [{\cal Q}_c \sum_{l=0}^\infty \bar G_{ss}^{\,l}]  \Gsr , \; 
{\cal Q}_c={\bf 1}-{\cal P}_c, \;
\bar G_{ss}={\cal Q}_c \Gss{\cal Q}_c .
\end{equation}

We characterize the strength of these 3 components by their respective weights $\Wrr$, $\Wpr$, $\Wqr$ given respectively by the sum of all matrix elements of  $\Grr$, $\Gpr$, $\Gqr$ divided by $N_r$. By definition we have $\Wrr + \Wpr + \Wqr =1$.

\section*{Results: $\GR$ properties}
\subsection*{Reduced Google matrix of country networks}

Reduced Google matrix has been computed, together with its components $\Grr$, $\Gpr$ and $\Gqr$, for the English language edition of Wikipedia (EnWiki) and for the 2 selected sets of 27 and 40 countries listed in Tables~\ref{tab:EU} and~\ref{tab:countries}. We recall that the 40 countries in the first set are the ones with top PageRank $K$ in the network of EnWiki. Countries are ordered by increasing $K$ value in all subsequent matrix representations.  The weight of the three matrix components of $\GR$ are listed in Table~\ref{tab:weight}. Predominant component is clearly $\Gpr$ but as we will explain next, it is not the most meaningful. 
\begin{table}[!ht]
\centering
\caption{{\bf Weights of the three matrices components of $\GR$}.}
\label{tab:weight}
\begin{tabular}{c|c|c|c|c|}
\cline{2-5}
                         & $\Wpr$  & $\Wqr$   & $\Wrr$   & Sum \\ \hline
\multicolumn{1}{|c|}{40} & 0.96120 & 0.029702 & 0.009098 & 1   \\ \hline
\multicolumn{1}{|c|}{EU} & 0.95332 & 0.038346 & 0.008334 & 1   \\ \hline
\end{tabular}
\end{table}

The meaning of $\Grr$ is clear as it is directly extracted from the global Google matrix $G$. It gives the direct links between the selected nodes and more specifically the probability $\Grr(i,j)$ for the surfer to go directly from column $j$ country to line $i$ country. 

\begin{figure}[!h]
\centering
\includegraphics[scale=0.7]{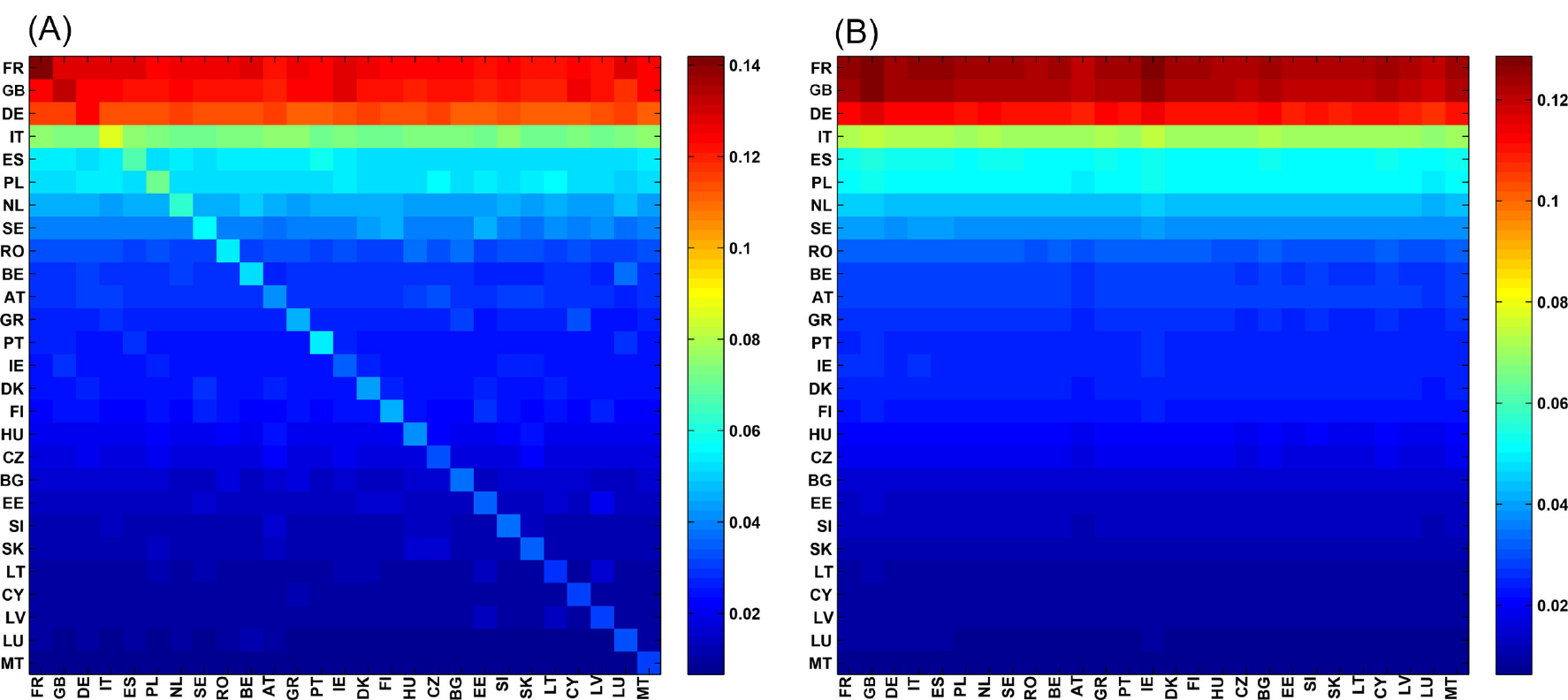}
\caption{{\bf Density plots of matrices $\GR$ and $\Gpr$ extracted from EnWiki for 27 EU countries. }
(A): $\GR$, (B): $\Gpr$. The colors represent maximum (red), intermediate (green) and minimum (blue) values.}
\label{Fig3}
\end{figure}

The sum of $\Gpr$ and $\Gqr$ represents the contribution of all indirect links through the scattering matrix $\Gss$. The projector component $\Gpr$ is rather close  to  nearly identical columns given by the PageRank probabilies of $N_r$ nodes (see Fig~\ref{Fig3}-(B)). Fig~\ref{Fig3} shows the matrix density plots for $\GR$ and $\Gpr$ for the 27 EU countries where lines and columns are ordered by increasing $K$ values. For both matrices, column values are proportional to their PageRank probabilities. As detailed in \cite{politwiki}, we observe numerically that $\Gpr \approx P_r\,E_r^T/(1-\lambda_c)$, meaning that each column is close to the normalized vector $P_r/(1-\lambda_c)$. As such, $\Gpr$ transposes essentially in $\GR$ the contribution of the first eigenvector of $G$. We can conclude that even if the overall column sums of $\Gpr$ account for $\sim 95$-97\% of the total column sum of $\GR$, $\Gpr$ doesn't offer innovative information compared to the legacy PageRank analysis. 

A way more interesting contribution is the one of $\Gqr$. This matrix captures higher-order indirect links between the $N_r$ nodes due to their interactions with the global network environment. We will refer to these links as \emph{hidden links}. 
We note that $\Gqr$ is composed of two parts $\Gqr = \Gqrd + \Gqrnd$ where the first diagonal term-only matrix $\Gqrd$ represents the probabilities to stay on the same node during multiple iterations of $\bar G_{ss}$ in (\ref{eq_3rdterm}) while the second matrix only captures non-diagonal terms in $\Gqrnd$. As such, $\Gqrnd$ represents indirect (hidden) links between the $N_r$ nodes appearing via the global network.
We note that a few matrix elements of $\Gqr$ can be negative, which is possible due to the negative terms in ${\cal Q}_c={\bf 1}-{\cal P}_c$ appearing in (\ref{eq_3rdterm}). The total weight of negative elements is however much smaller than $\Wqr$ (at least 6 times smaller and even non-existing in ArWiki for the network of 40 countries).

\begin{figure}[!ht]
\centering
\includegraphics[scale=0.7]{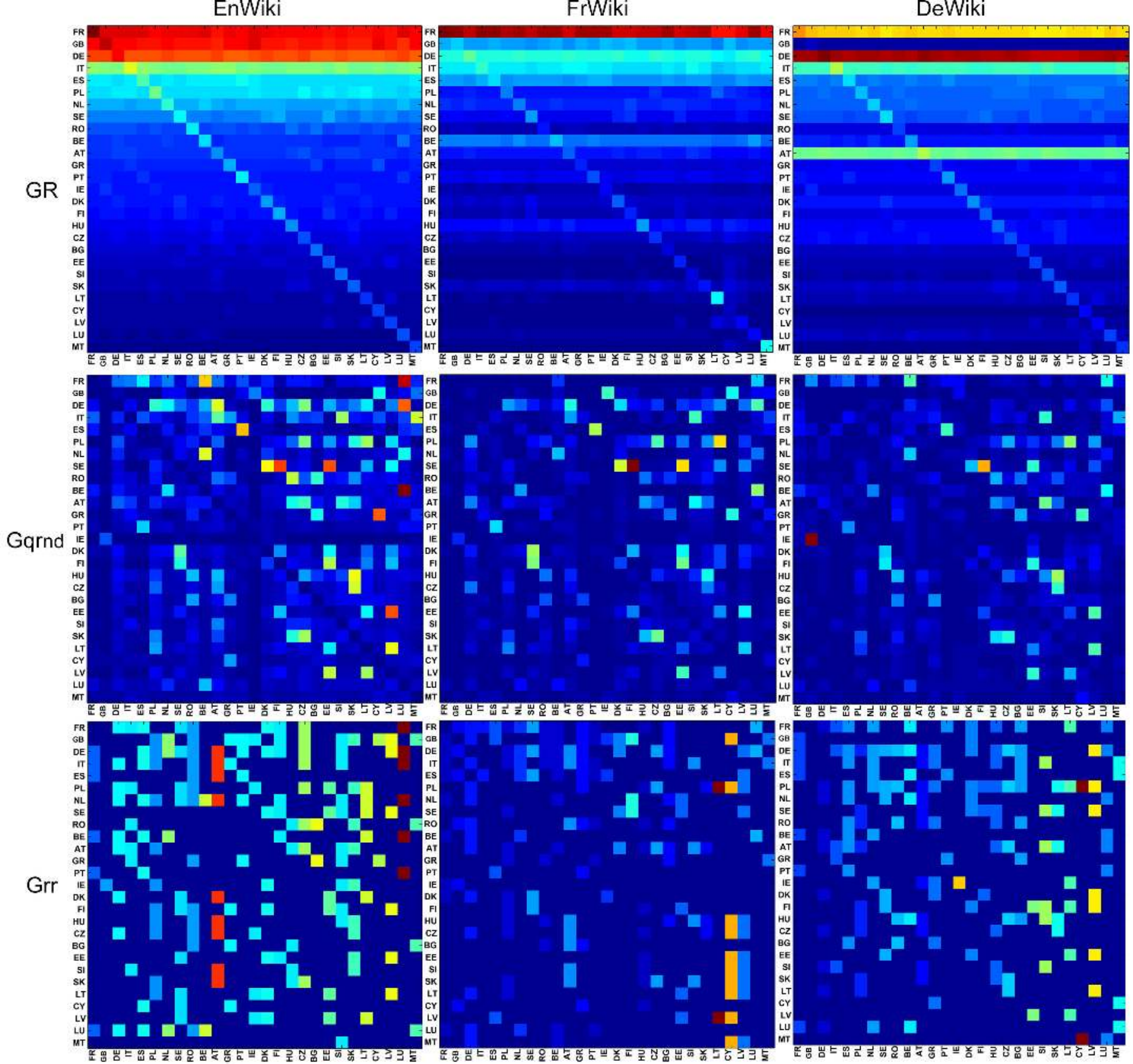}
\caption{{\bf Density plots of $\GR$, $\Gqrnd$ and $\Grr$.}
$\GR$ (first line), $\Gqrnd$ (second line) and $\Grr$ (third line) for the reduced network of EU countries of EnWiki (left column), FrWiki (middle column) and DeWiki (right column).  
The nodes $N_r$ are ordered in lines and columns by the reference PageRank of EnWiki. The colors represent maximum (red), intermediate (green) and minimum (blue) values.}
\label{Fig4}
\end{figure}

For the three EnWiki, FrWiki and DeWiki editions, Fig~\ref{Fig4} plots the density of matrices $\GR$, $\Gqrnd$ and $\Grr$. We keep for all plots the same order of countries extracted from the EnWiki network. This is meant to highlight cultural differences among Wikipedia editions. From the first line of Fig~\ref{Fig4}, it is clear that  $\GR$ matrix is dominated by the projector $\Gpr$ contribution, which is proportional to the global PageRank probabilities. 
Several cultural biases can be extracted from $\GR$. For instance, France is the top country in EnWiki and FrWiki, while Germany is the top country in DeWiki. 

The information from hidden links between countries is provided by $\Gqrnd$. 
It shows, for the three selected languages editions, the strong hidden links connecting Finland to Sweden. Other interesting hidden links are between Ireland and United Kingdom in DeWiki or in EnWiki linking Luxembourg to France.
The reduced Google matrix density plots for the network of 40 worldwide countries are to be found in reference \cite{geop}.

\FloatBarrier
\subsection*{Networks of friends and followers}

As proposed in \cite{geop}, it is possible to extract from $\GR$ and $\Gqrnd$ a network of friends and followers to easily illustrate direct and hidden links in the network. Direct links are given by $\Grr$ while hidden (i.~e.~ indirect) are given by $\Gqrnd$. For the sake of simplicity, we refer next to $\Gqrnd$ using $\Gqr$ notation. 

\begin{figure}[!ht]
\centering
\includegraphics[scale=0.7]{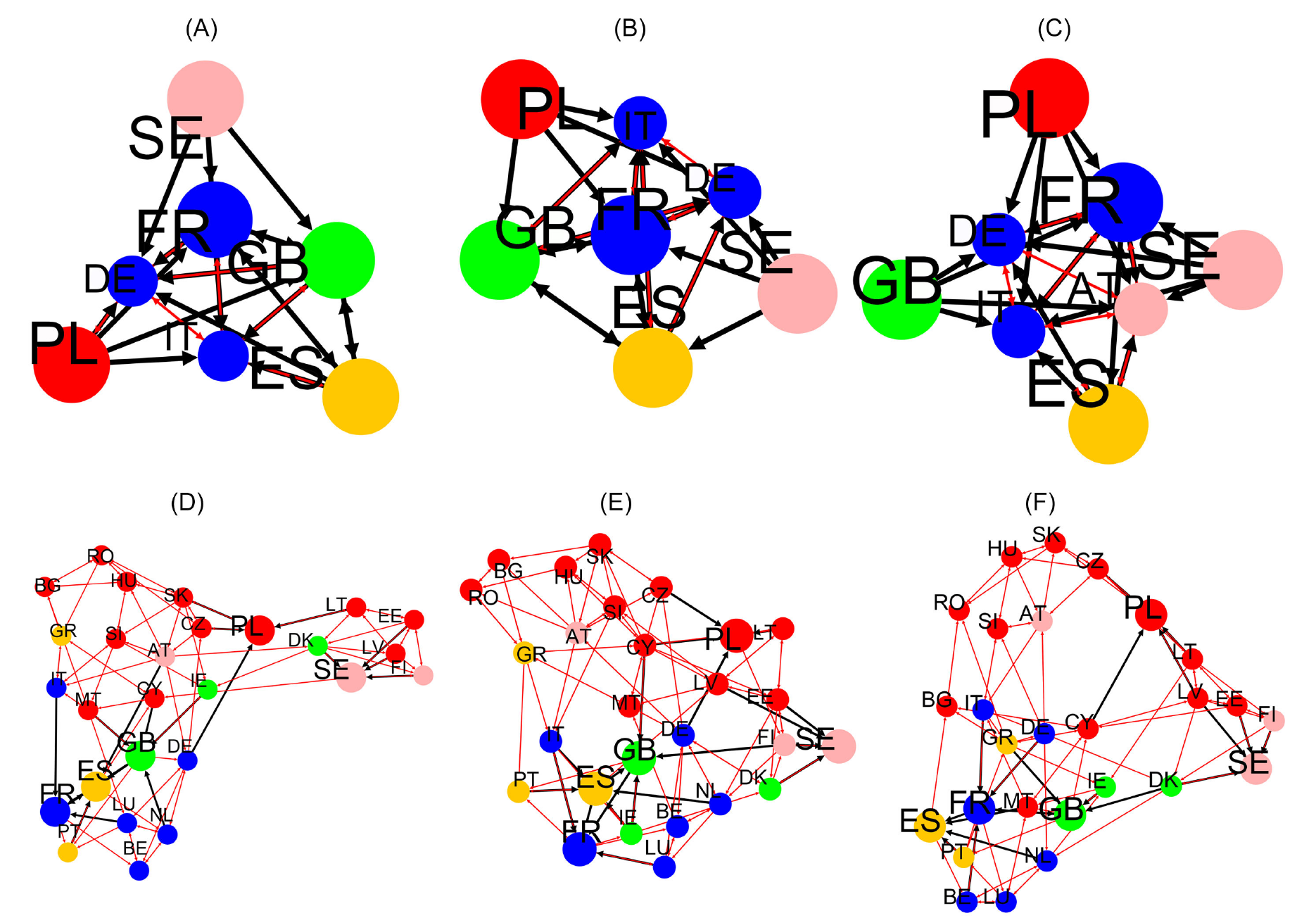}
\caption{{\bf Relationships structure extracted from $\GR$ for the network of EU countries.}
friends (top line) and followers (bottom line) induced by the 5 top countries of each group (FR, GB, ES, SE, PL). Results are plotted for EnWiki (A and D), FrWiki (B and E) and DeWiki (C and F). Node colors represent geographic appartenance to a group of countries (cf. Fig~\ref{Fig1} and Table~\ref{tab:EU} for details).
Top (bottom) graphs: a country node with higher PageRank probability has a bigger size and points (is pointed by) with a bold black arrow to its top 4 friends (followers). Red arrows show friends of friends (resp. followers of followers) interactions computed until no new edges are added to the graph.}
\label{Fig5}
\end{figure}

To create these networks of friends and followers, we divide the set of $N_r$ nodes into representative groups as shown in Fig~\ref{Fig1} for 27 EU country set. EU countries are grouped upon their accession date to the union (e.g. Founder, 1973, 1981-1986, 1995, 2004-2007). One leading country per EU member state group has been selected as follows:
 \begin{itemize}
 	 \item France for Founders,
 	 \item United Kingdom for countries having joined in 1973,
 	 \item Spain for countries having joined between 1981 and 1986,
 	 \item Sweden for countries having joined in 1995,
 	 \item Poland for countries having joined between 2004 and 2007.
 \end{itemize}

For each leading country $j$, we extract from both matrices $\Gqr$ and $\GR$ the top 4 \emph{Friends} (resp. \emph{Followers}) given by the 4 best values of the elements of column $j$ (resp. of line $j$). In other words, it corresponds to destinations of the 4 strongest outgoing links of $j$ and the countries at the origin of the 4 strongest ingoing links of $j$. These networks of top 4 friends and followers have been calculated for the five editions of Wikipedia. 

\begin{figure}[!ht]
\centering
\includegraphics[scale=0.7]{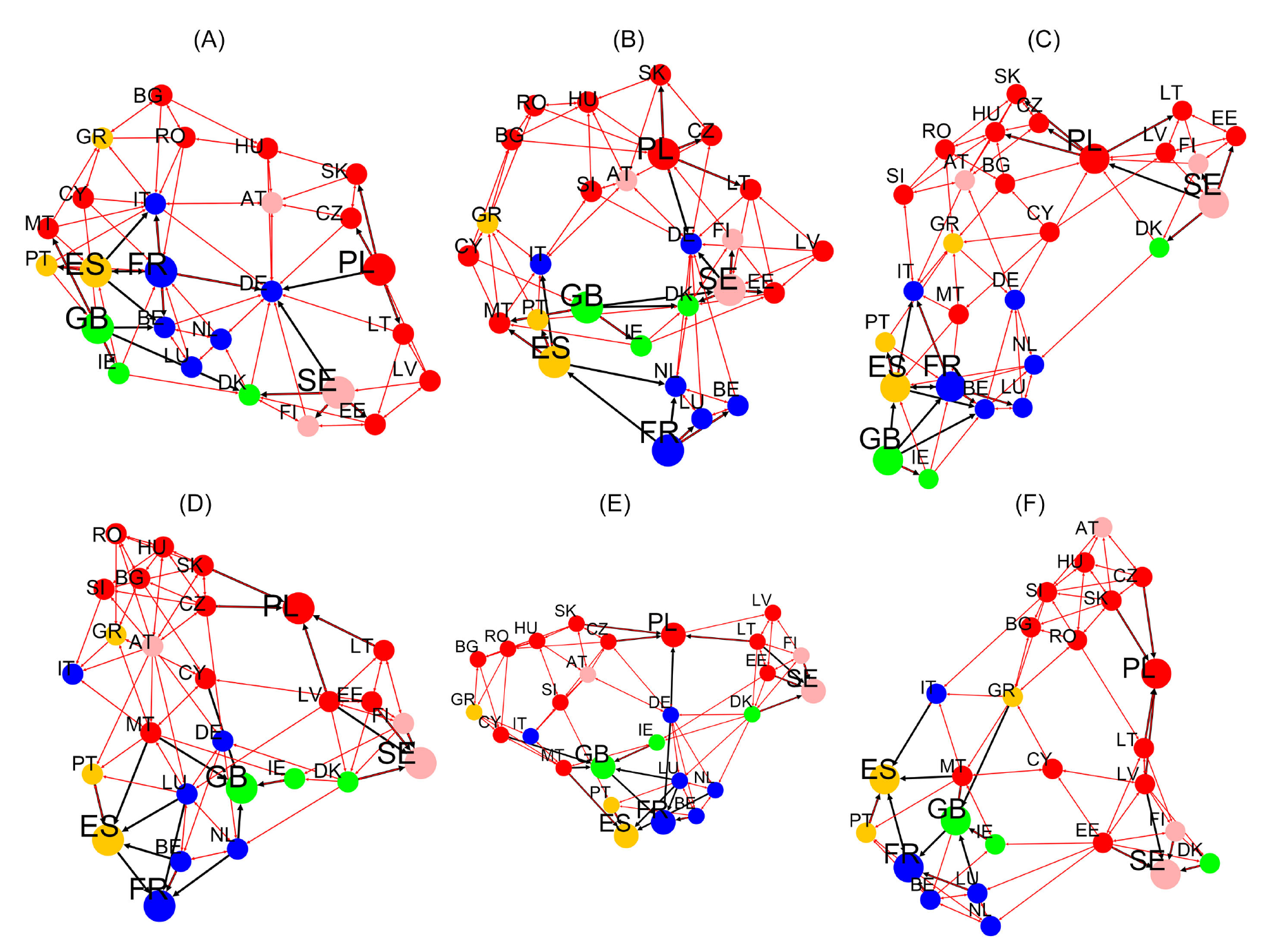}
\caption{{\bf Relationship structure extracted from $\Gqrnd$ for the network of EU countries.}
Same legend as Fig~\ref{Fig5} except friends and followers are computed from $\Gqrnd$.}
\label{Fig6}
\end{figure}

Top 4 friends and top 4 followers of EU leading countries are extracted from $\GR$ and $\Gqr$ to plot the graphs of Fig~\ref{Fig5} and~\ref{Fig6}. Results for EnWiki, FrWiki and DeWiki are presented here. Note that Fig~\ref{Fig6} pictures hidden links. The black thick arrows identify the top 4 friends and top 4 followers interactions. Red arrows represent the friends of friends (respectively the followers of followers) interactions that are computed recursively until no new edge is added to the graph. All graphs are visualized with the Yifan Hu  layout algorithm~\cite{YifanHu} using Gephi \cite{gephi}.



The vertices of the network of friends obtained from $\GR$ concentrate, for each Wiki, to about 7 countries, 5 of which being the leading ones. The other vertices are top PageRank countries such as Italy, Germany or Spain. This is due to the predominance of PageRank probabilities in the structure of $\GR$. 
A more valuable information could be extracted from the network of followers. In all editions,  Benelux and Nordic countries create a cluster densely interconnected. The networks of followers end up spanning the full set of EU countries in this representation. On this representation, it can be noticed that the order of arrival of member states is meaningful. Indeed, nodes of the same color are closely interconnected.  

The hidden friends and followers relationships are extracted from $\Gqr$ and illustrated in Fig~\ref{Fig6}. 
As discussed earlier, $\Gqr$ is not dominated by PageRank, and as such, the resulting network of friends includes more nodes and shows more diversity. 
It is worth noting that Germany, as one of the Founders, bridges the group of Founders to Sweden (the leader of the countries that have joined EU in 1995) and Poland (the leader of the countries that have joined EU between 2004 and 2007) in FrWiki and EnWiki. From EnWiki and DeWiki, strong ties are seen between Italy and France, while it is not the case from FrWiki authors. This is another example of cultural bias. However, lots of links are to seen in all three editions: GB-IE, SE-FI, ES-PT, PL-LT, IT-GR and many others. To underline this constant presence of links, we give in Table~\ref{tab:euGqr} the list of friends (resp. followers) that are among the top 4 ones in all 5 editions, in 4 out of 5 and in 3 out of 5 for $\Gqr$ analysis. For each leading country, around 2 to 3 top friends and followers exist accros all editions. 

For the 40 worldwide countries set, networks of top 4 friends and followers are to be found in \cite{geop}, calculated for the same 5 editions of Wikipedia as well. Similar observations have been made as for the set of 27 EU countries. 


\begin{table}[!ht]
\begin{adjustwidth}{-2.25in}{0in}
\centering
\caption{{\bf Cross-edition friends and followers extracted from $\Gqr$ of EU countries} per leading country.}
\label{tab:euGqr}
\begin{tabular}{|c @{\extracolsep{\fill}} l |l | l || l |l |l|} 
\hline
          Top  & \multicolumn{3}{c||}{$G_{\rm qr}$ Wiki friends present in} & \multicolumn{3}{c|}{$G_{\rm qr}$ Wiki followers present in} \\ \cline{2-7}
country &  all 5 editions     & 4 out of 5 editions       & 3 out of 5 editions   & all 5 editions     & 4 out of 5 editions       & 3 out of 5 editions    \\ \hline
FR      & BE -ES	& IT	&			& BE		& LU - ES	&                   \\
GB      & IE		& 	& DK - FR		& IE - MT	& CY		&                   \\
ES      & IT - PT	& FR	& BE			& MT - PT	& 		& LU                  \\
SE      & DK - FI	& 	& EE			& DK - EE - FI	& LV		&                   \\
PL      & CZ		& 	& DE - HU - LT - SK	& CZ - LT - SK	& 		& LV     \\ \hline
\end{tabular}
\begin{flushleft}
For each top country, we list the friends and followers that are identical accros all five Wikipedia editions, in 4 editions out of 5 and in 3 editions out of 5.
\end{flushleft}
\end{adjustwidth}
\end{table}

\FloatBarrier
\section*{Results: $GR$ link sensitivity}
\subsection*{Influence analysis of geopolitical ties using $\GR$}

We have now established the global mathematical structure $\GR$ and presented how it can be leveraged to extract meaningful geopolitical interactions among countries for the two sets of interest, naming 27 EU and 40 worldwide countries. 
These interactions are extracted from Wikipedia and thus stem from all links covering this very rich network of webpages.  As such, they encompass not only interactions related to economics or politics, but from any possible domain (arts, history, entertainment, etc.). The strength of this study is to show that just from the structure of the network, relevant and timely information can be extracted. The hyperlinked structure of Wikipedia itself contains an important part of the universal knowledge stored in details on the webpages.  

Previous study has shown that $\GR$ captures essential interactions between countries. The point is now to see how some ties between countries influence the whole network structure. More specifically, we focus here on capturing the impact of a change in the strength of a relationship between two countries on the importance of the nodes in the network. Therefore we have designed a sensitivity analysis that measures a logarithmic derivative of the PageRank probability when the transition probability of only one link is increased for a specific couple of nodes in $\GR$, relatively to the other ones.

Our sensitivity analysis is performed for a directed link where the relationship going from country $i$ to $j$ is increased.  We investigate in the last part of this Section the imbalance between the influence of two opposite direction interactions. In other words, we conduct the aforementioned sensitivity analysis for the link going from country $i$ to $j$, and for the link going in the opposite direction from $j$ to $i$. For each pair of countries, we derive from this two-way sensitivity the relationship imbalance to identify the most important player in the relationship.  

\subsection*{Sensitivity analysis}
We define $\delta$ as the relative fraction to be added to the relationship from nation $j$ to nation $i$ in $\GR$. 
Knowing $\delta$, a new modified matrix $\GRprime$ is calculated in two steps. First, element $\GRprime(i,j)$ is set to $(1+\delta)\cdot\GR(i,j)$. Second, all elements of column $j$ of $\GRprime$ are normalized to 1 (including element $i$) to preserve the unity column-normalization property of the Google matrix.
Now $\GRprime$  reflects an increased probability for going from nation $j$ to nation $i$.

It is now possible to calculate the modified PageRank eigenvector $\Pprime$ from $\GRprime$ using the standard $\GRprime\Pprime = \Pprime$ relation and compare it to the original PageRank probabilities $P$ calculated with $\GR$ using $\GR P = P$. The same process can be applied to the transposed version of $\GRprime$ to calculate the modified CheiRank probabilities $\tilde{P}^*$.
Due to the relative change of the transition probability between nodes $i$ and $j$, steady state PageRank and CheiRank probabilities are modified. This reflects a structural modification of the network and entails a change of importance of nodes in the network. 
These changes are measured by a logarithmic derivative of the PageRank probability of node $a$:

\begin{equation}
D_{(j \rightarrow i)}(a) = {({\rm d}P_a}/{{\rm d} \delta_{ij})/P_a} = {(\Pprime_a - P_a)}/({\delta_{ij}}{P_a}) 
\label{eq_sensitivity}
\end{equation}
Notation ${(j \rightarrow i)}$ indicates that the link from node $j$ to node $i$ has been modified. 
Element $D_{(j \rightarrow i)}(a)$ gives the logarithmic variation of PageRank probability for country $a$ if the link from $j$ to $i$ has been modified. We will refer to this variation as the \emph{sensitivity} of nation $a$ to the relationship from nation $i$ to nation $j$. If this sensitivity is negative, country $i$ has lost importance in the network. On the opposite, a positive sensitivity expresses a gain in importance.  
The computation has been tested for values of $\delta = \pm 0.01, \pm 0.03, \pm 0.05$. The result is not sensitive to $\delta$ and following results are given for $\delta=0.03$. 


\subsection*{Relationship imbalance analysis}

As introduced earlier, sensitivity $D_{(j \rightarrow i)}(k)$ of Eq~\eqref{eq_sensitivity} measures the change of importance of node $a$ if the link from nation $j$ to $i$ has been changed. The sensitivity of node $a$ to a change in one direction is not necessarily the same as its sensitivity to the change in the opposite direction. We define as such the \emph{2-way sensitivity} of node $a$ which is simply the sum of the sensitivities calculated for both directions: 
\begin{equation} 
D_{(i \leftrightarrow j)}(a) = D_{(i \rightarrow j)}(a) + D_{(j \rightarrow i)}(a)
\label{ep_2way}
\end{equation}

The two-way sensitivity can be leveraged to find out, for a pair of countries $a$ and $b$, which one has the most influence on the other one. Therefore, we define the following metric : 
 \begin{equation}
\label{eq_F}
F(a,b) = D_{(a \leftrightarrow b)}(a) - D_{(a \leftrightarrow b)}(b)
\end{equation}
Here, we measure the 2-way sensitivity for nodes $a$ and $b$ when the link between them is modified both ways in $\GR$.
If $F(a,b)$ is positive, it means that the 2-way sensitivity of $a$ is larger than the 2-way sensitivity of $b$. In this case, $a$ is more influenced by $b$ than $b$ by $a$. We can say that $b$ is the \emph{strongest} country. 
If $F(a,b)$ is negative, we can say that $a$ is the strongest country.  



\section*{Sensitivity results}\label{sec:results}

Sensitivity analysis results are shown first for the 27 EU network and then for the 40 worldwide network. For each network, we have identified a set of meaningful links between countries to be modified and observed resulting sensitivity of other nations. We perform as well for each network the relationship imbalance analysis for each pair of nations. 
Note that if the modified link is clearly identified, we will drop the index $i \rightarrow j$ in our sensitivity measure notation for clarity. 

\subsection*{27 EU network of countries}

In order to better capture the countries' sensitivities from a multicultural perspective, we have calculated the sensitivities for 3 Wikipedia editions: EnWiki, FrWiki and DeWiki. All sensitivity results shown for 27 EU network have been averaged over the three editions as follows:
\begin{equation}
\bar{D}=\frac{1}{3}\sum_{i=1}^3 D^i
\end{equation}where index $i$ refers to the Wikipedia edition. 

\subsubsection*{Sensitivity analysis}
%

We start this analysis by introducing a first simple example where Italy increases its relationship with France. Then, we analyze the impact on the EU countries of Great Britain's exit (i.e. Brexit) from European Union. Next, we highlight the sensitivity of Luxembourg to the increase of Germany and France's cooperation with other member states. Finally, we present the results that underline the strong ties that exist between groups of countries that function together in Europe.

\begin{figure}[!ht]
\centering
\includegraphics[scale=0.5]{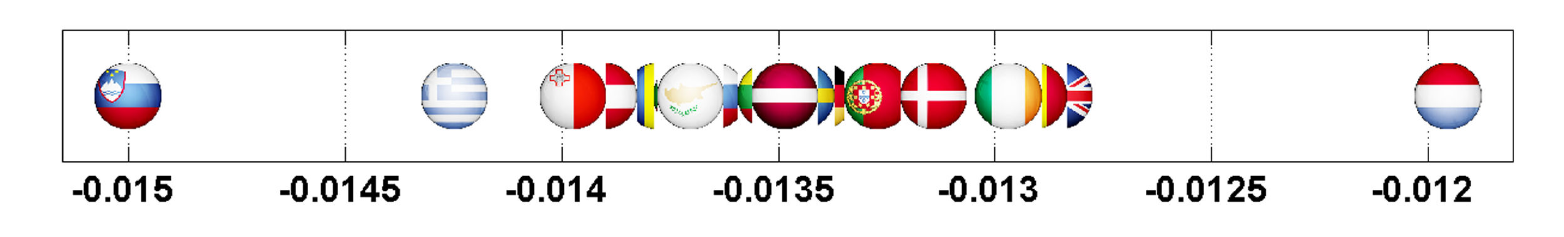}
\caption{{\bf Axial representation of $\bar{D}$ for a link modification from \{IT\} to \{FR\}.}
Here $\bar{D}(IT)=-0.0159$ and $\bar{D}(FR)=0.0701$ are not shown.}
\label{Fig7}
\end{figure}

\begin{figure}[!ht]
\centering
\includegraphics[scale=0.5]{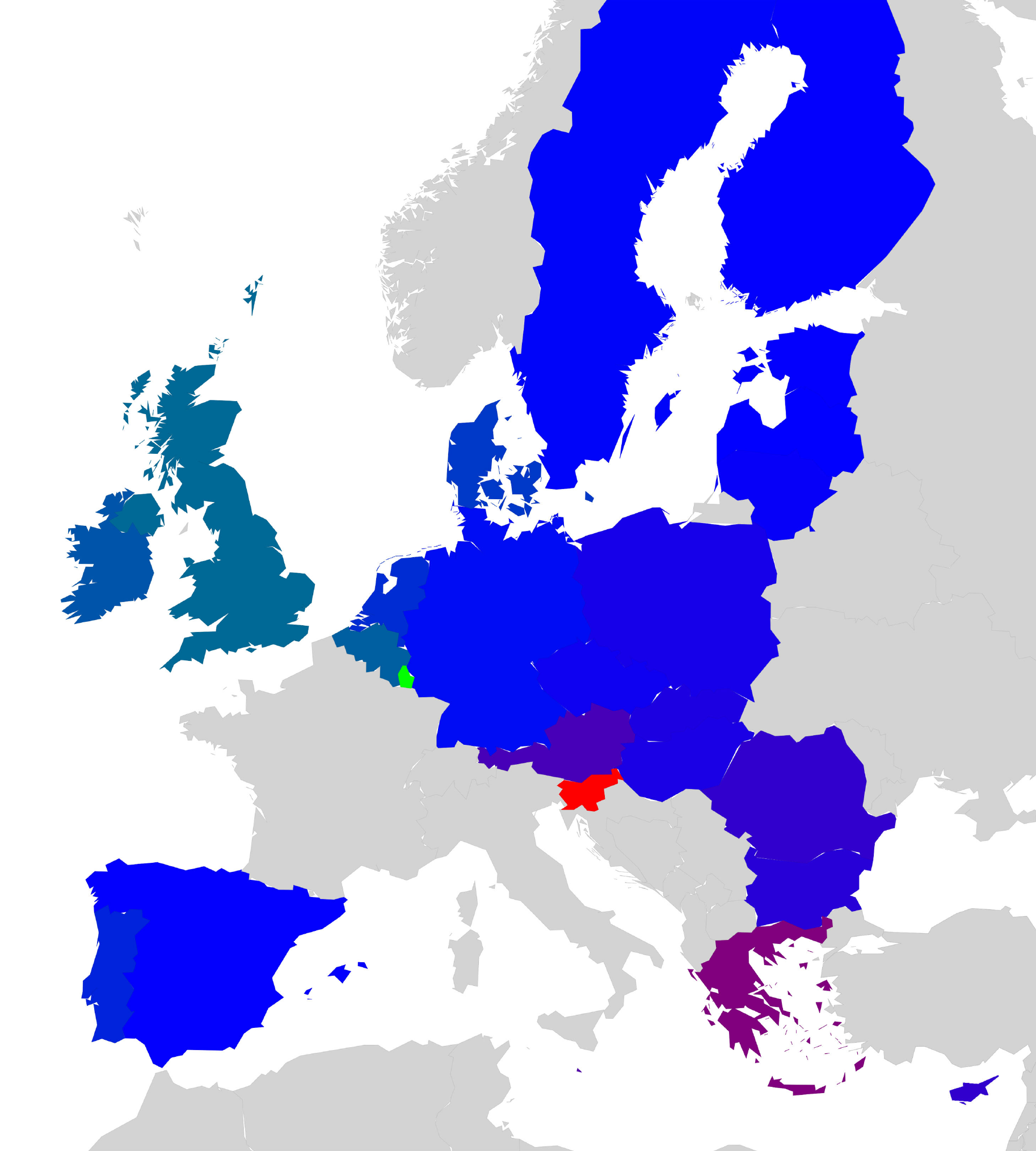}
\caption{{\bf Map representation of $\bar{D}$ for links modifications from \{IT\} to \{FR\}. }
Lower values of $\bar{D}$ in red, median in green and larger in blue (IT and FR are not shown).
}
\label{Fig8}
\end{figure}

\begin{figure}[!ht]
\centering
\includegraphics[scale=0.8]{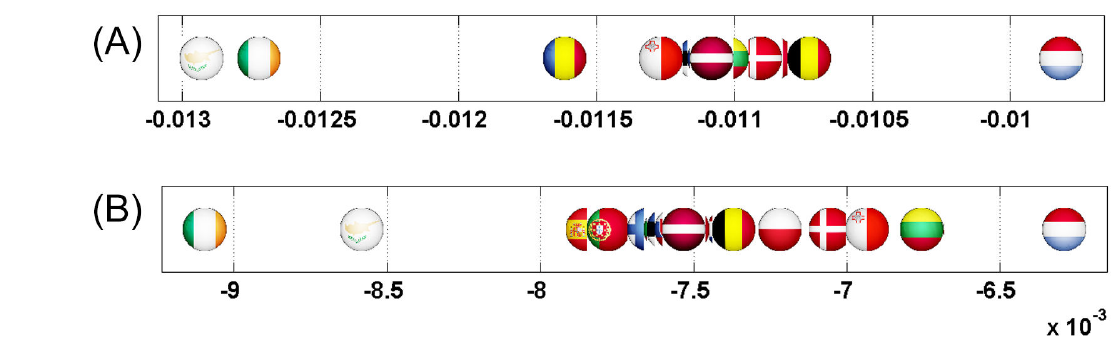}
\caption{{\bf Axial representation of $\bar{D}$ for link modifications from \{GB\} to \{FR or DE\}.}
(A): GB to FR (not shown $\bar{D}(GB)=-0.0124$ and $\bar{D}(FR)=0.0577$). (B): GB to DE 
(not shown $\bar{D}(GB)=-0.0087$ and $\bar{D}(DE)=0.0606$).}
\label{Fig9}
\end{figure}

\begin{figure}[!ht]
\centering
\includegraphics[scale=0.5]{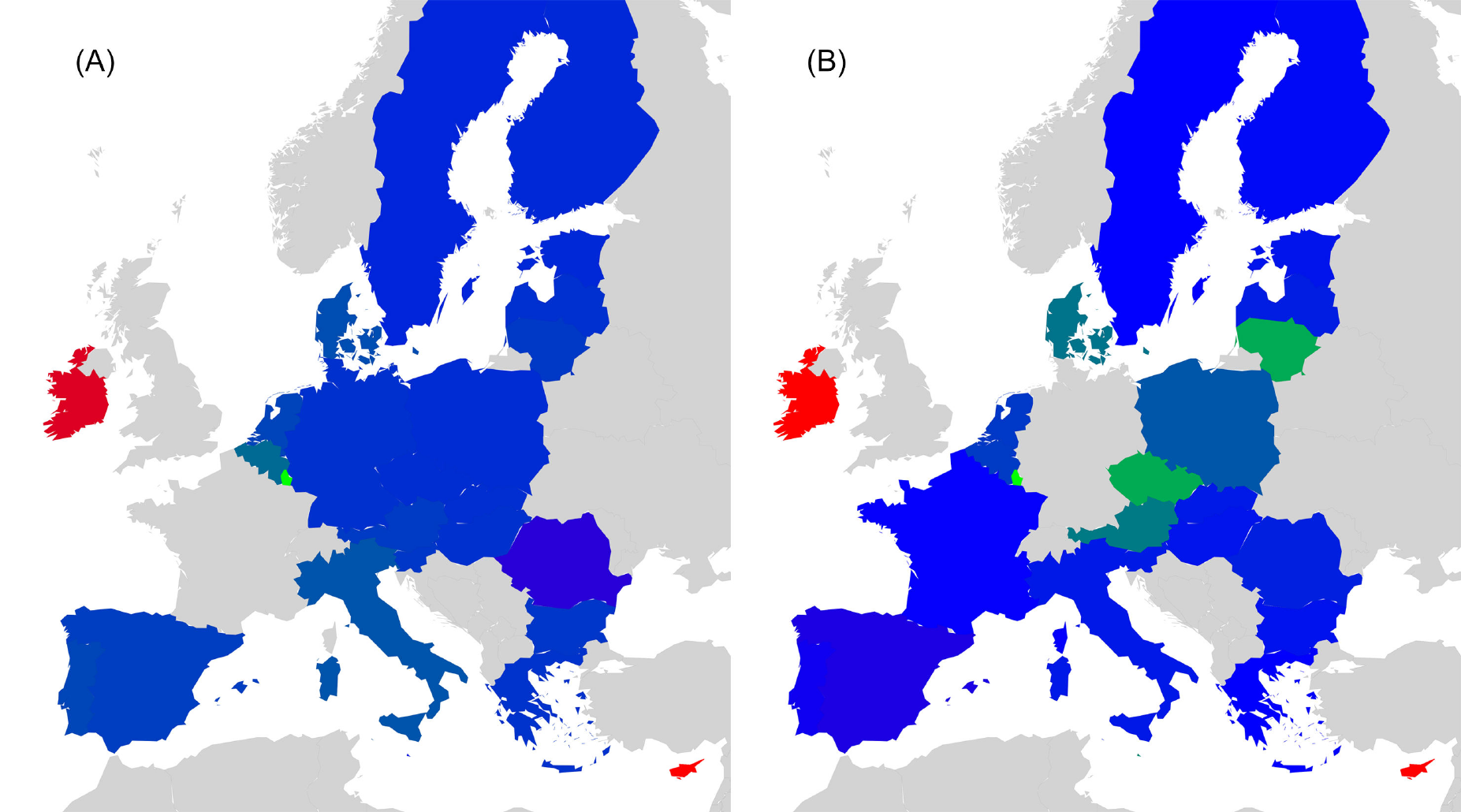}
\caption{{\bf Map representation of $\bar{D}$ for link modifications from \{GB\} to \{FR or DE\}. }
(A): GB to FR ($\bar{D}$ are not shown for GR, FR); 
(B): GB to DE($\bar{D}$ are not shown for GR, DE). 
Lower values of $\bar{D}$ in red, median in green and larger in blue.}
\label{Fig10}
\end{figure}

\begin{figure}[!ht]
\centering
\includegraphics[scale=0.7]{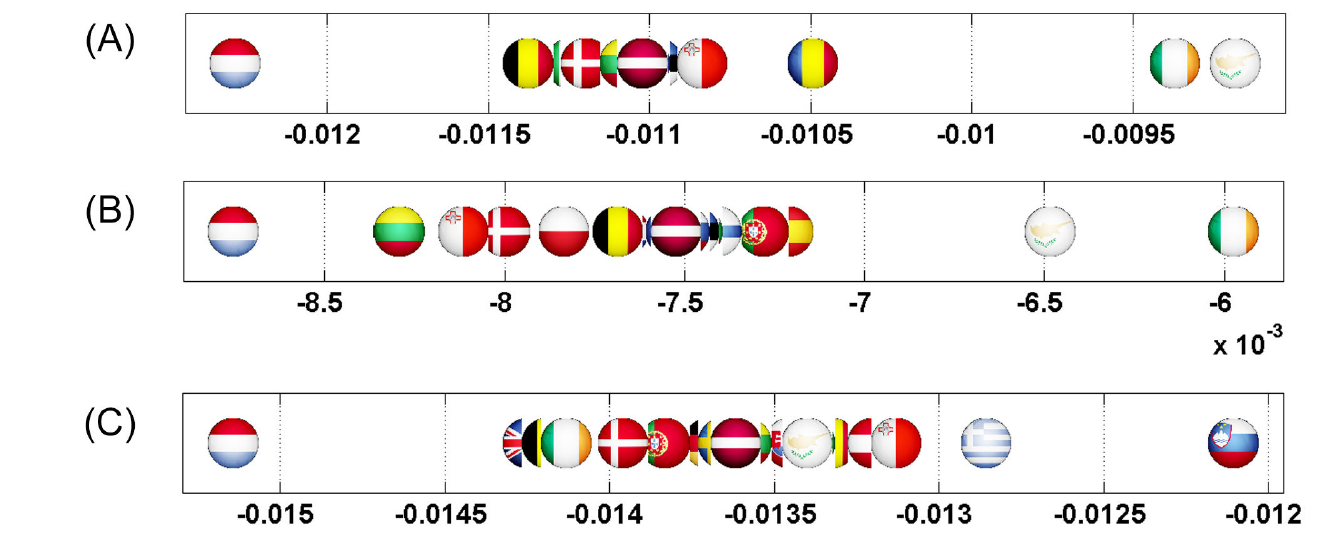}
\caption{{\bf Axial representation of $\bar{D}$ for link modifications from \{FR or DE\} to \{GB or IT\}.}
(A): FR to GB (not shown $\bar{D}(FR)=-0.0117$ and $\bar{D}(GB)=0.1572$). 
(B): DE to GB (not shown $\bar{D}(DE)=-0.0081$ and $\bar{D}(GB)=0.1248$). 
(C) FR to IT (not shown $\bar{D}(FR)=-0.0143$ and $\bar{D}(IT)=0.1508$).}
\label{Fig11}
\end{figure}

\begin{figure}[!ht]
\centering
\includegraphics[scale=0.7]{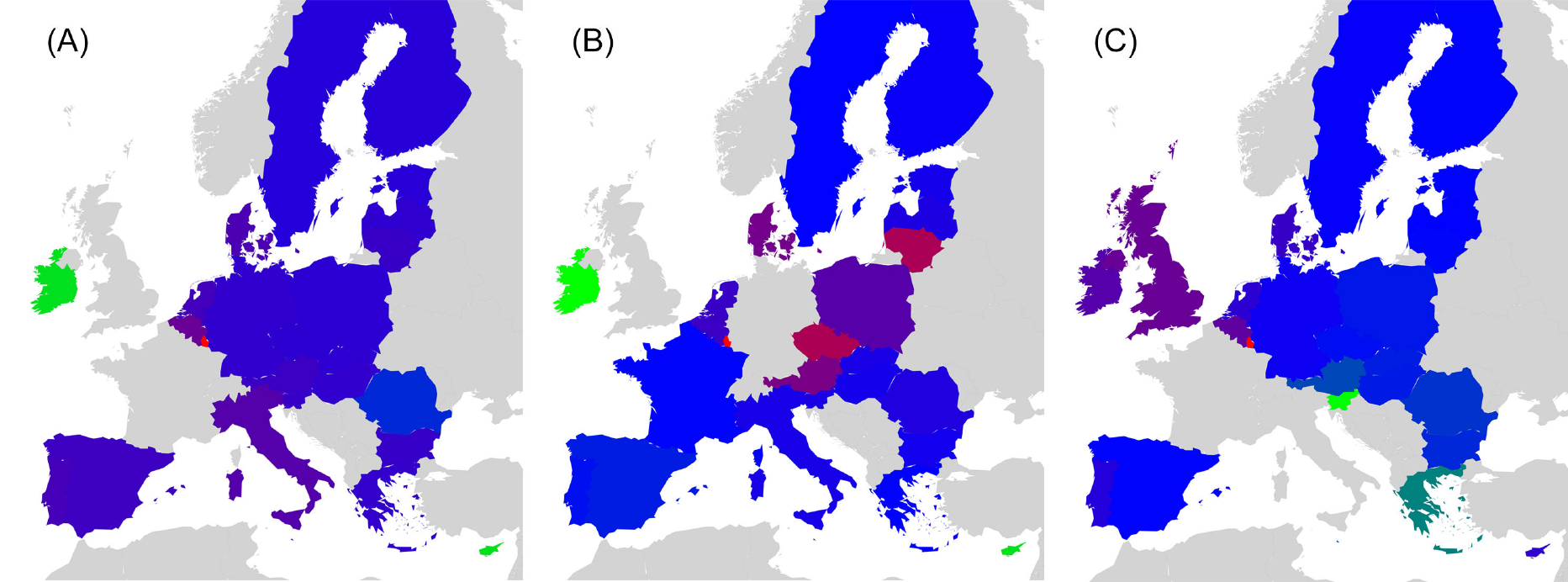}
\caption{{\bf Map representation of $\bar{D}$ for link modifications from \{FR or DE\} to \{IT or GB\} : Luxembourg is negatively impacted here. }
(A): FR to GB. (B): DE to GB. (C) FR to IT. 
Lower values of $\bar{D}$ in red, median in green and larger in blue; for linked countries $\bar{D}$ is not shown.}
\label{Fig12}
\end{figure}

\begin{figure}[!h]
\centering
\includegraphics{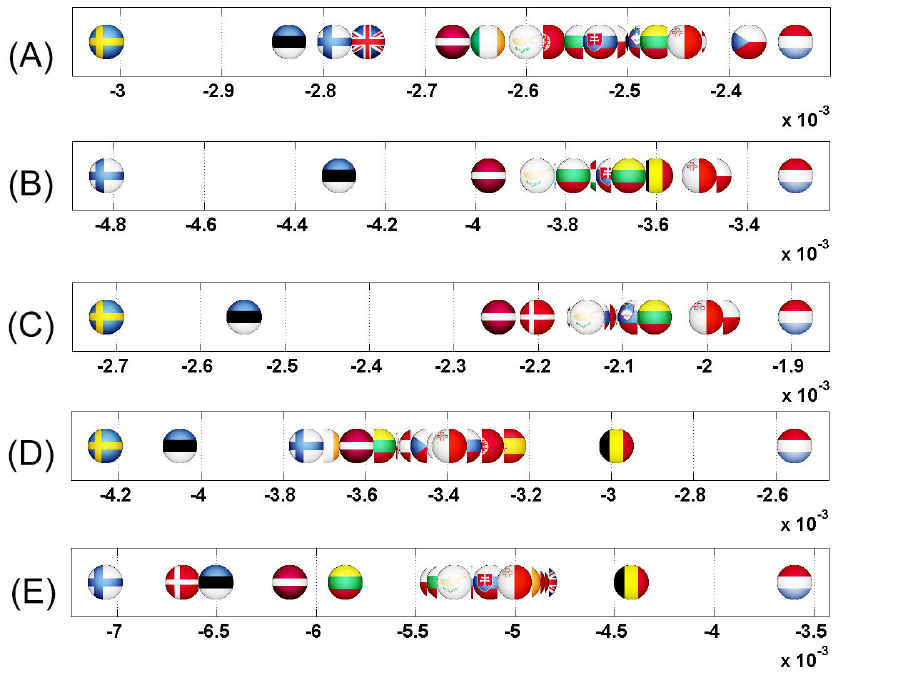}
\caption{{\bf Axial representation of $\bar{D}$ for link modifications from Nordic countries to \{FR or DE\}.}
(A): DK to DE (not shown $\bar{D}(DK)=-0.0050$ and $\bar{D}(DE)=0.0208$). 
(B): SE to DE (not shown $\bar{D}(SE)=-0.0064$ and $\bar{D}(DE)=0.0313$). 
(C): FI to DE (not shown $\bar{D}(FI)=-0.0046$ and $\bar{D}(DE)=0.0173$). 
(D): DK to FR (not shown $\bar{D}(DK)=-0.0077$ and $\bar{D}(FR)=0.0197$). 
(E): SE to FR (not shown $\bar{D}(SE)=-0.0100$ and $\bar{D}(FR)=0.0296$).}
\label{Fig13}
\end{figure}

\begin{figure}[!h]
\centering
\includegraphics[scale=0.7]{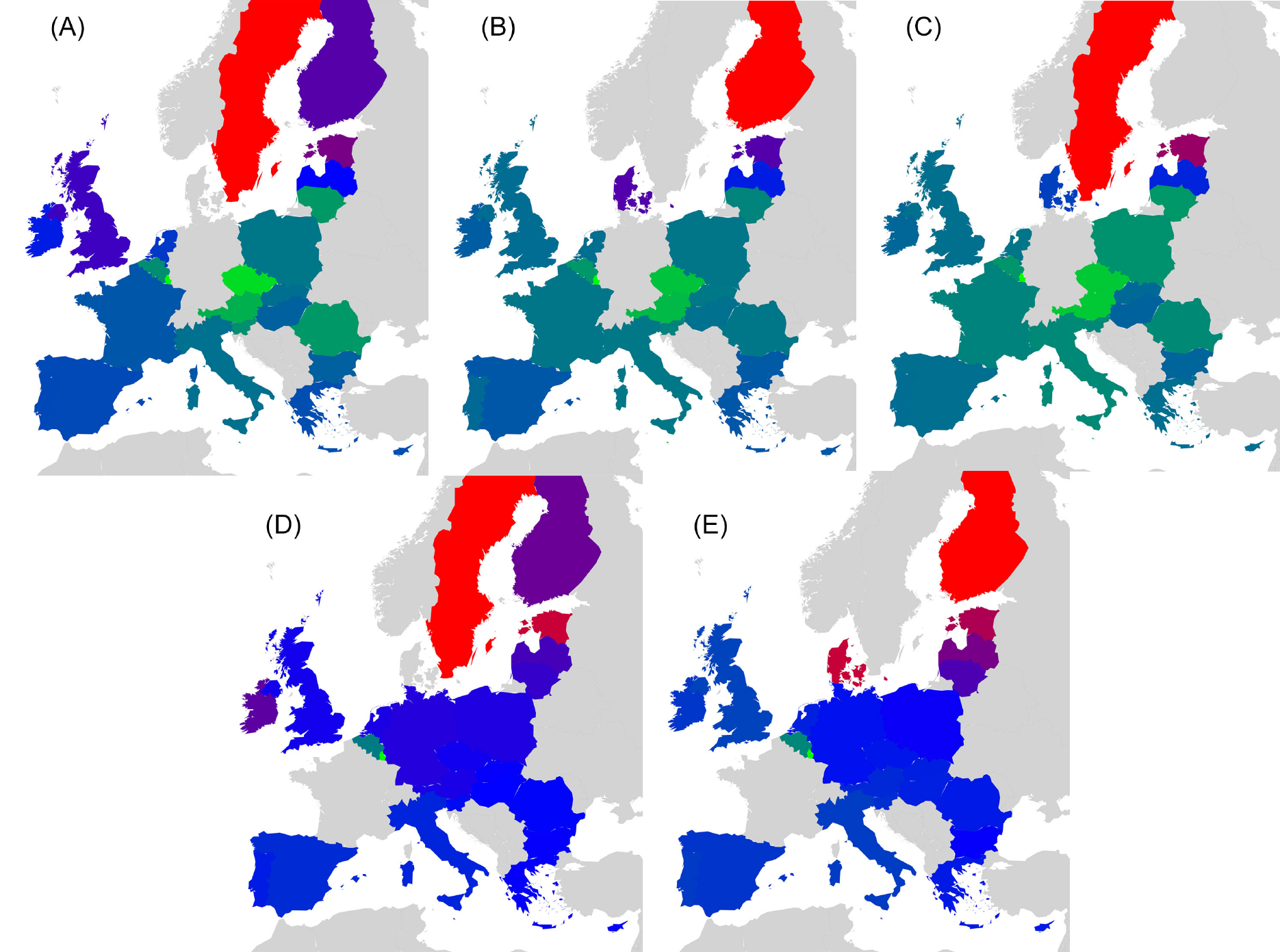}
\caption{{\bf Map representation of $\bar{D}$ for link modifications from Nordic countries to \{FR or DE\}.}
(A): DK to DE. (B): SE to DE. (C): FI to DE. (D): DK to FR. (E): SE to FR. Lower values of $\bar{D}$ in red, median in green and larger in blue; for linked countries $\bar{D}$ is not shown.}
\label{Fig14}
\end{figure}

\begin{figure}[!h]
\centering
\includegraphics{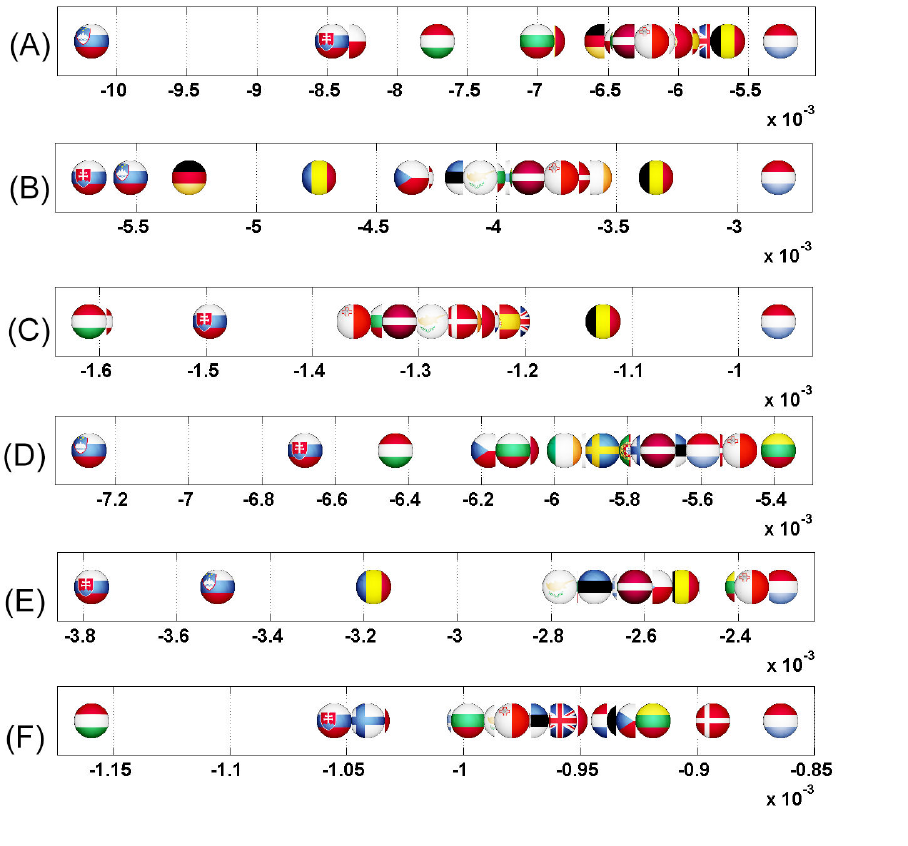}
\caption{{\bf Axial representation of $\bar{D}$ for link modifications from \{AT, HU and SI\} to \{FR or DE\}.}
(A): AT to FR (not shown $\bar{D}(AT)=-0.0101$ and $\bar{D}(FR)=0.0373$). 
(B): HU to FR (not shown $\bar{D}(HU)=-0.0080$ and $\bar{D}(FR)=0.0205$). 
(C): SI to FR (not shown $\bar{D}(SI)=-0.0046$ and $\bar{D}(FR)=0.0075$). 
(D): AT to DE (not shown $\bar{D}(AT)=-0.0070$ and $\bar{D}(DE)=0.0393$). 
(E): HU to DE (not shown $\bar{D}(HU)=-0.0052$ and $\bar{D}(DE)=0.0311$). 
(F): SI to DE (not shown $\bar{D}(SI)=-0.0034$ and $\bar{D}(DE)=0.0081$).}
\label{Fig15}
\end{figure}

\begin{figure}[!h]
\centering
\includegraphics[scale=0.7]{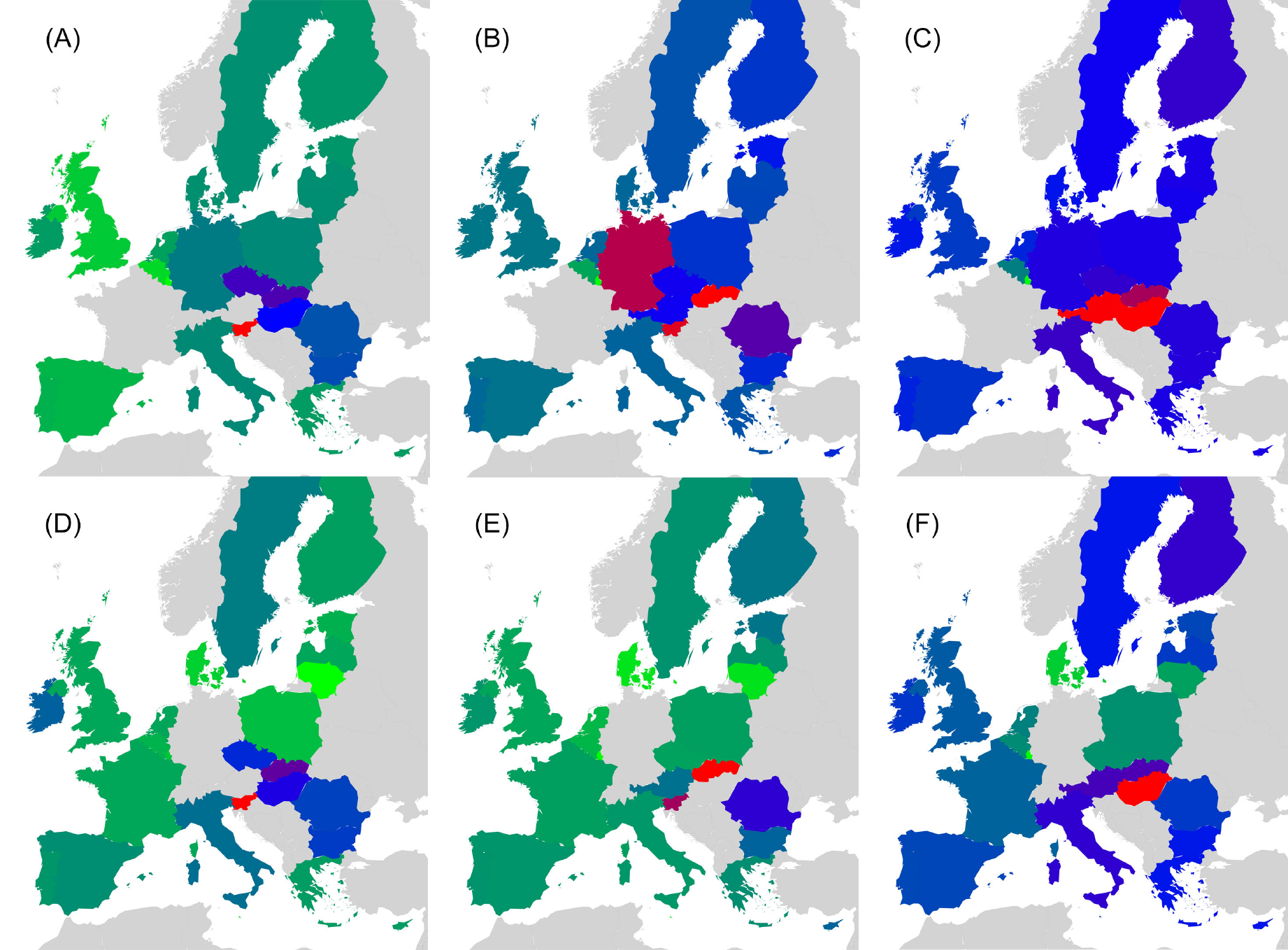}
\caption{{\bf Map representation of $\bar{D}$ for link modifications from \{AT, HU and SI\} to \{FR or DE\}.}
(A): AT to FR. (B): HU to FR. (C): SI to FR. (D): AT to DE. (E): HU to DE. (F): SI to DE. Lower values of $\bar{D}$ in red, median in green and larger in blue; for linked countries $\bar{D}$ is not shown.}
\label{Fig16}
\end{figure}

For each sensitivity analysis, we show two types of figures: $i)$ an axial representation of the sensitivity $\bar{D}$ (cf. Fig~\ref{Fig7}, Fig~\ref{Fig13}, Fig~\ref{Fig15}, Fig~\ref{Fig9}, Fig~\ref{Fig11}) and $ii)$ a colored map of Europe where countries' color indicate the sensitivity $\bar{D}$ as well (cf. Fig~\ref{Fig8}, Fig~\ref{Fig14}, Fig~\ref{Fig16}, Fig~\ref{Fig10}, Fig~\ref{Fig12}). Color scale for these maps plots lower values of $\bar{D}$ in red, median in green and larger in blue.  Each map represents the sensitivity values obtained for a given link variation. 

\paragraph{Italy to France relationship}
Italy is the second top export and import country of Slovenia with \$3.05B and \$3.84B respectively. In 1992, diplomatic relations began between the two countries and in 2012, Foreign Minister of Italy, Giulio Terzi, described the bilateral  relationship between Italy and Slovenia as fruitful and dynamic~\cite{italy}. Politically, Slovenia relies on Italy to become a member of the principal UN, EU and NATO bodies~\cite{italy}. No doubt Slovenia would suffer if Italy decided to go away from it and increase its relationships with France. The 27 EU network exactly shows the negative impact of Italy increasing its link in $\GR$ with France: Slovenia is the nation with lowest sensitivity on Fig~\ref{Fig7} and~\ref{Fig8}.

\FloatBarrier

\paragraph*{Impact of Brexit\footnote{Brexit is an abbreviation for Britain exit~\cite{bbcbrexit}.}}

The United Kingdom has triggered article 50 on March 27, 2017 to leave the European Union as a consequence of the referendum of June 23rd, 2016~\cite{bbcbrexit}. To understand its impact on EU countries with our dataset, we have reduced (and not increased as done in other studies) the $\GR$ transition probability UK towards France or Germany. We remind that our network is dated by 2013 but it captures the strong UK influence. Results are shown in Fig~\ref{Fig9} and~\ref{Fig10} and indicate that Ireland and Cyprus are by far the most negatively affected countries in both cases. Moreover, the sensitivity of UK is negative as it benefits less from France's or Germany's influence.  
These facts have been recently backed up by specialists. In~\cite{brexit}, a study delivered by the London School of Economics discussing the consequences of Brexit forecasts that UK will loose $2.8\%$ of its GDP\footnote{Gross domestic product (GDP) is the monetary value of all the finished goods and services produced within a country's borders in a specific time period~\cite{investopedia}.}. Similarly, \cite{brexit} shows that Ireland will loose as well $2.3\%$ of its GDP, which is the largest proportional loss caused by Brexit. Cyprus-UK Relations are strong as claimed by the official website of the Ministry of Foreign Affairs of Cyprus~\cite{cyprus}. Referring to~\cite{atlas}, UK is the $4^{th}$ top export destination for Cyprus with \$242M and the $2^{nd}$ import origin with \$508M. As such, this clear bond of UK with Cyprus explains that if GB suffers from Brexit, Cyprus will do as well. Our data strikingly exhibits the same conclusion as shown in Fig~\ref{Fig9} and~\ref{Fig10}.

\FloatBarrier
\paragraph*{Luxembourg's sensitivity to Germany and France}

Luxembourg shares its borders with Belgium, Germany and France with whom it has strong and diverse relationships. Luxembourg has a very open economy. Together with Belgium, they position themselves as the $12^{th}$ largest economy in the world. Two of the top three export and import countries of Belgium-Luxembourg are Germany (\$44.6B, \$50.4B) and France (\$43.8B, \$36.8B)~\cite{atlas}. Official languages in Luxembourg are Luxembourgish, French and German. Luxembourg has robust relationships with France~\cite{luxfr,luxfr2} and Germany~\cite{luxde} in various areas such as finance, culture, science, security or nuclear power.
It is clear that Luxembourg will suffer if one of these European countries reduces its exchanges with it. In Fig~\ref{Fig11} and~\ref{Fig12}, we clearly show with our sensitivity analysis that Luxembourg is strongly influenced by France and Germany. If France or Germany increases its relationships with Italy or Great Britain, Luxembourg is by far the most negatively impacted country.

\FloatBarrier

\FloatBarrier

\paragraph*{Clusters of countries}

By analyzing the sensitivity of countries to various 2-nation relationships, we have noticed that several groups of nations function together. These groups are strongly interconnected, and if anyone of these group members increases its relationship strength with a country outside of the group, all group members loose importance in the network. We highlight  two meaningful examples next: the cluster of Nordic countries and the cluster Austro-Hungarian cluster. Other clusters we have identified in our network are for instance the cluster of Benelux countries (e.g. Belgium, the Netherlands and Luxembourg) or the cluster of the Iberian peninsula (e.g. Portugal and Spain). 

For both investigated groups, we test the influence of an increase in collaboration from one member of the group to France or to Germany. France and Germany have been chosen as they are central members of European Union.

The Nordic countries Denmark, Finland, and Sweden have much in common: their way of life, history, language and social structure ~\cite{nordic}. After World War II, the first concrete step into unity was the introduction of a Nordic Passport Union in 1952. Nordic countries co-operate in the Nordic Council, a geopolitical forum. In the Nordic Statistical Yearbook ~\cite{nordic}, Klaus Munch illustrates that ``The Nordic economies are among the countries in the Western World with the best macroeconomic performance in the recent ten years''. Nordic countries should keep cooperating to stay strong. Thus, if any Nordic country attempts to abandon these relationships in favor of other countries, it will negatively impact the remaining Nordic countries. Our sensitivity analysis illustrates this impact in Fig~\ref{Fig13} and~\ref{Fig14}. In these figures, we show how the relationship increase between any Nordic country towards France or Germany induces a drop in sensitivity for Nordic countries.

Referring to~\cite{slovenia}, relations between Slovenia, Hungary and Austria are tight. Hungary has supported Slovenia for its NATO membership applications and Austria has assisted Slovenia in entering European Union. Relationships between Austria and Hungary are important for both countries in the economic, political and cultural fields~\cite{athu}. Concerning economy~\cite{atlas}, Austria is one of the top import origins for Hungary and Slovenia with \$5.54B and \$2.37B respectively. Similarly to the Nordic group of countries, if Austria, Slovenia or Hungary increases its relationships with another European country, the other two will be affected. Sensitivity analysis backs up this statement as seen in Fig~\ref{Fig15} and~\ref{Fig16}.  

\FloatBarrier

\subsubsection*{Relationship imbalance analysis}

\begin{figure}[!ht]
\centering
\includegraphics[scale=0.9]{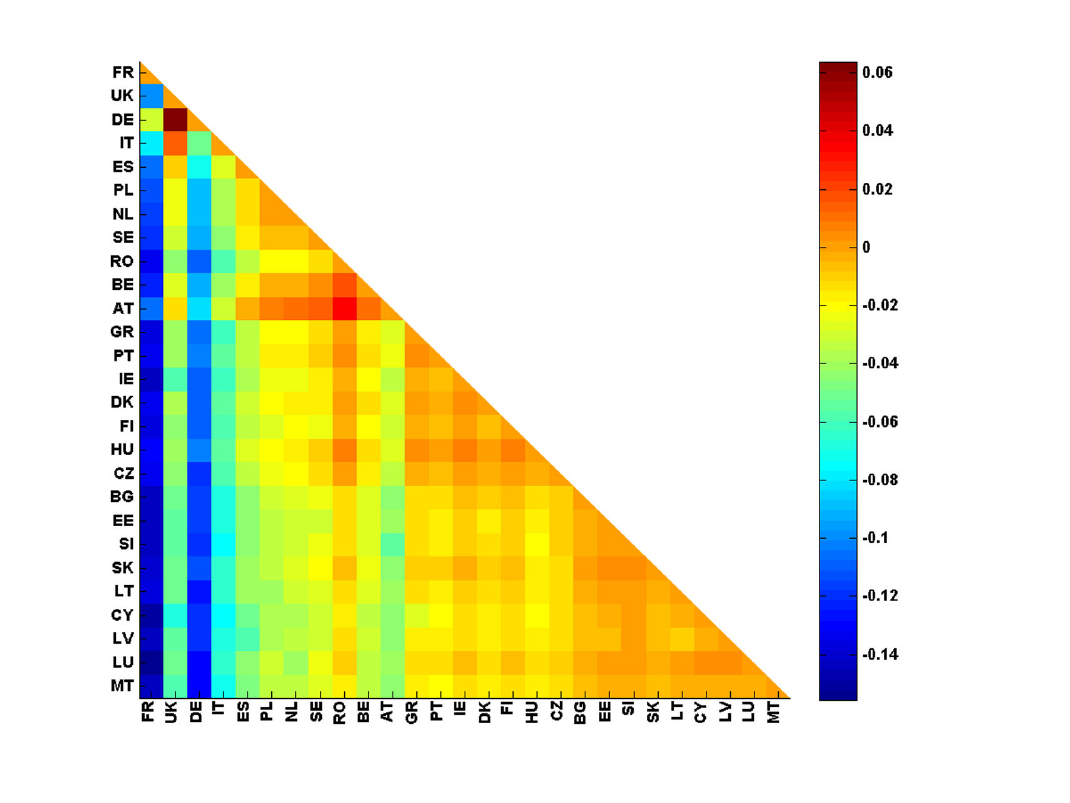}
\caption{{\bf Relationship imbalance analysis: F-representation for 27 EU network.} 
$F(a,b)$ is given by the colorbar. X-axis and Y-axis represent $a$ and $b$ respectively. If $F(a,b)$ is negative, nation $a$ has more influence on nation $b$ than $b$ on $a$.  
}
\label{Fig17}
\end{figure}

Relationship imbalance analysis has been derived for all pairs of European countries following Eq~\eqref{eq_F}.
Fig~\ref{Fig17} shows a density plot of $F(a,b)$. 
We recall that if $F(a,b)$ is negative, nation $a$ has more influence on nation $b$ than $b$ on $a$. If $F(a,b)$ is positive, nation $b$ dominates nation $a$.
According to {\it The Globe of Economic Complexity} \cite{globe} and identical to our results in Fig~\ref{Fig17}, Germany and France are the two largest economies in Europe. From $\GR$ we can clearly see the dominance of France and Germany on other EU countries. Another interesting result of Fig~\ref{Fig17} is the equal influence between all pairs of countries created by one member of \{GR, PT, IE, DK, FI, HU\} and another of \{BG, EE, SI, SK, LT, CY, LV, LU, MT\}. These pairs have $F(a,b)$ close to zero and are plotted with orange color in Fig~\ref{Fig17}.

\FloatBarrier
\subsection*{40 worldwide network of countries}

Similarly to the 27 EU countries dataset, sensitivity results are averaged over 5 Wikipedia editions:  ArWiki, EnWiki, FrWiki, RuWiki and DeWiki. We first show as well the sensitivity analysis for carefully selected links and then conclude this part with the sensitivity imbalance analysis for all pairs of countries. 

\subsubsection*{Sensitivity Analysis}

In this worldwide set of countries, we have identified relationships whose impact on the network clearly shows how meaningful the sensitivity analysis proposed in this paper is. 

\begin{figure}[!ht]
\centering
\includegraphics{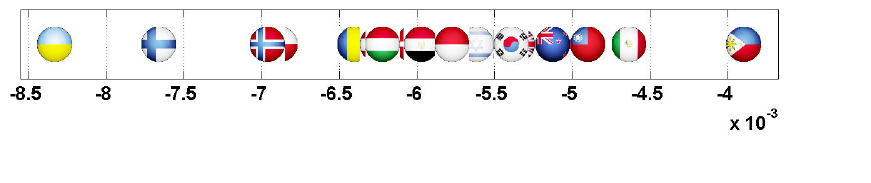}
\caption{{\bf Axial representation of $\bar{D}$ for link modification from RU to US.} 
(not shown $\bar{D}(RU)=-0.0089$ and $\bar{D}(US)=0.0446$).}
\label{Fig18}
\end{figure}

\begin{figure}[!ht]
\centering
\includegraphics[scale=0.7]{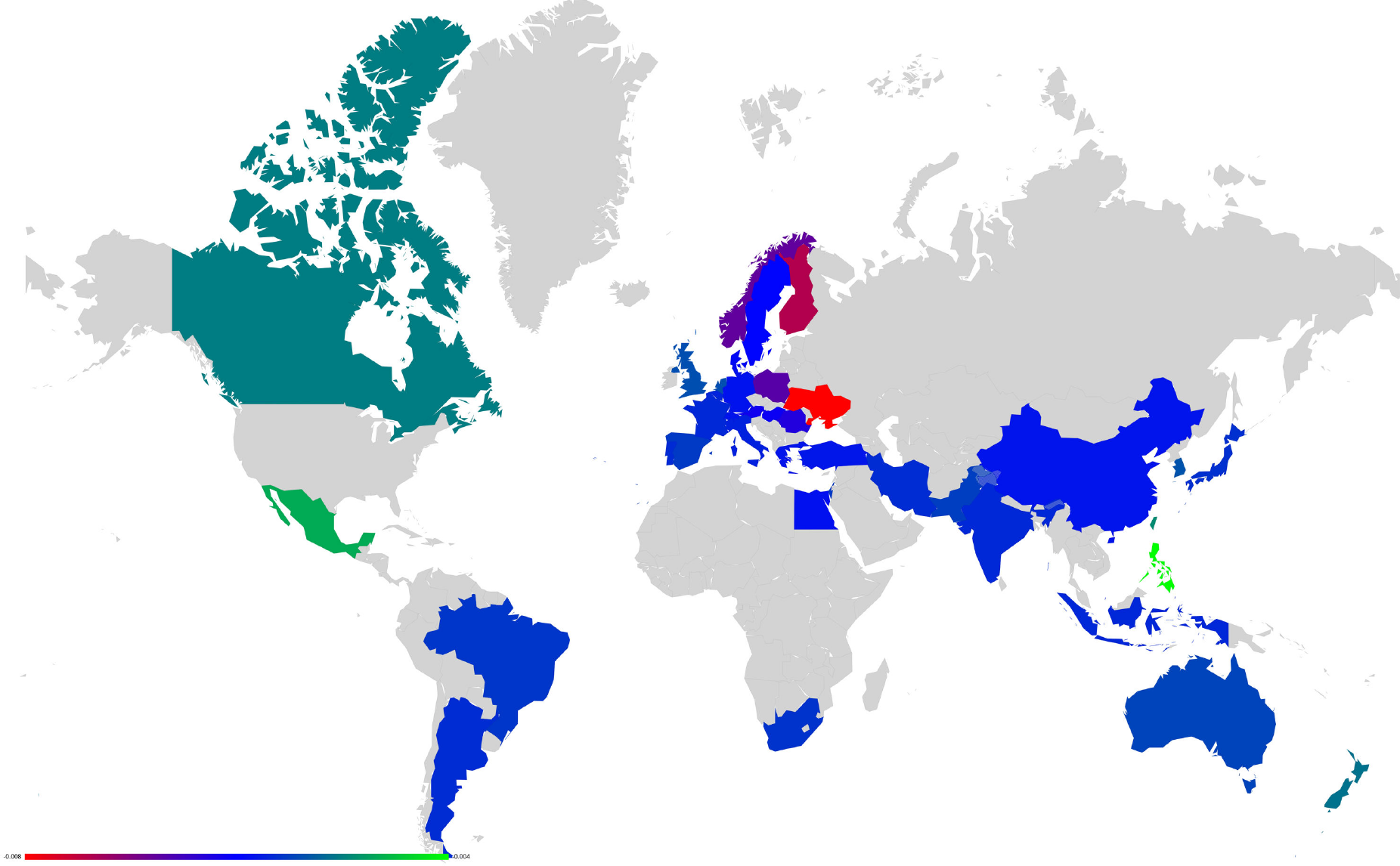}
\caption{{\bf Map representation of $\bar{D}$ for link modification from RU to US.}
For linked countries  $\bar{D}$ is not shown.
}
\label{Fig19}
\end{figure}

\paragraph*{US - Russia.}
As mentioned previously in the introduction, and according to the results in Fig~\ref{Fig18} and~\ref{Fig19},  Ukraine would be the most affected country if Russia gets closer to US. 
This is due to the fact that Ukraine and Russia were both in the USSR and their economies are 
strongly interconnected. The next influenced country is Finland which also has strong economic relations 
with Russia being a part of Russian Empire till beginning of 20th century.

\FloatBarrier

\begin{figure}[!ht]
\centering
\includegraphics[scale=0.7]{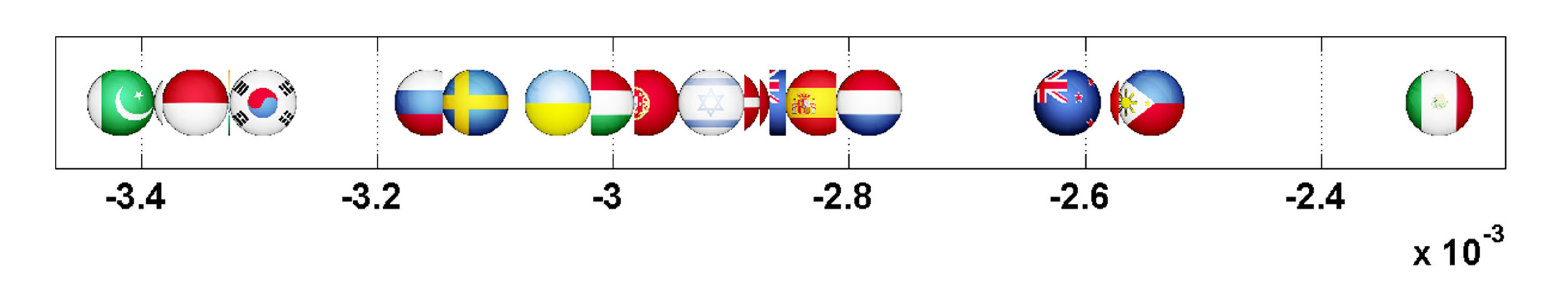}
\caption{{\bf Axial representation of $\bar{D}$ for link modification from CN to US (discarding TW).}
(not shown $\bar{D}(CN)=-0.0056$, $\bar{D}(US)=0.0210$ and $\bar{D}(TW)=-0.0087$).}
\label{Fig20}
\end{figure}

\begin{figure}[!ht]
\centering
\includegraphics[scale=0.7]{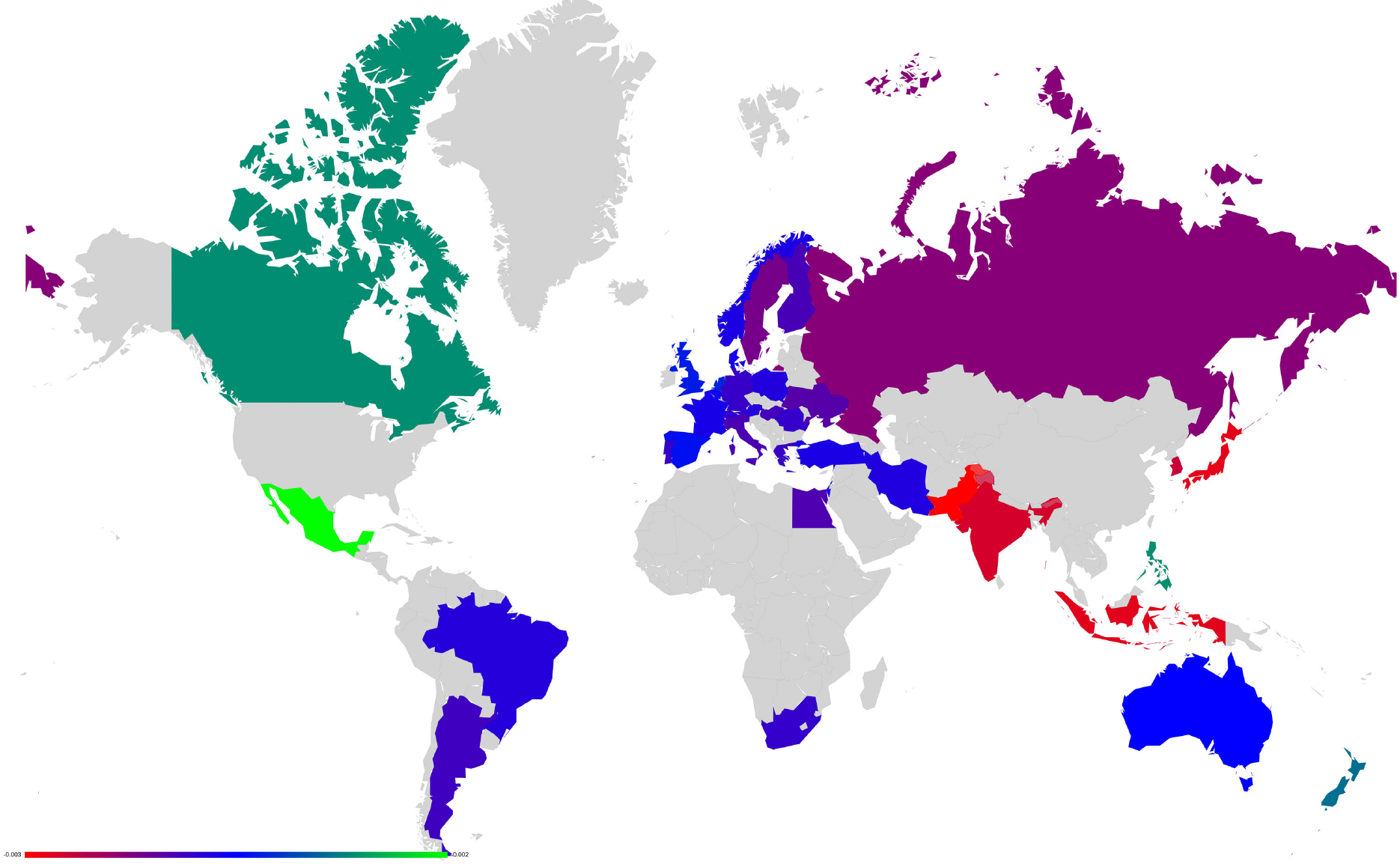}
\caption{{\bf Map representation of $\bar{D}$ for link modification from CN to US (discarding TW).}
For linked countries  $\bar{D}$ is not shown.}
\label{Fig21}
\end{figure}

 \paragraph*{China to US.} The effects of an increase in the relationship from China to US are shown in Fig~\ref{Fig21} and~\ref{Fig20}. Taiwan and Pakistan are the most negatively affected countries. Taiwan is not pictures in Fig~\ref{Fig21} and~\ref{Fig20} as it greatly reduces readability of the plots. Indeed, sensitivity of Taiwan is $\bar{D}(TW)=-0.0087$, 4 times the one of Pakistan. BBC's article~\cite{bbc} on the division between China and Taiwan illustrates that US is the most important friend and the only ally of Taiwan. China claims Taiwan as its territory and Taiwan counts on US to establish its full independence to stand up against China. As such, if the ties between China and US get stronger, Taiwan will loose its best ally. 

In 1951, Pakistan and China officially established their diplomatic relations and in 2016 they celebrated 65 years of friendship~\cite{pakistan}. Regarding security strategy, China has always supported Pakistan in facing terrorism. Politically, Pakistan stands with China on many issues concerning China's core interests (e.g. Taiwan, Tibet, Xinjiang). The trade volume between the two countries reached \$100.11B by 2015 and in 2016 the \$46B China-Pakistan Economic Corridor (CPEC)~\cite{cpec} was constructed. If China strengthens its relationship with US, Pakistan may clearly suffer from it. An article by Ian Price~\cite{sabotage} raises a serious question on whether United States aims at sabotaging the CPEC in the near future.

%

\FloatBarrier
 
\begin{figure}[!ht]
\centering
\includegraphics[scale=0.7]{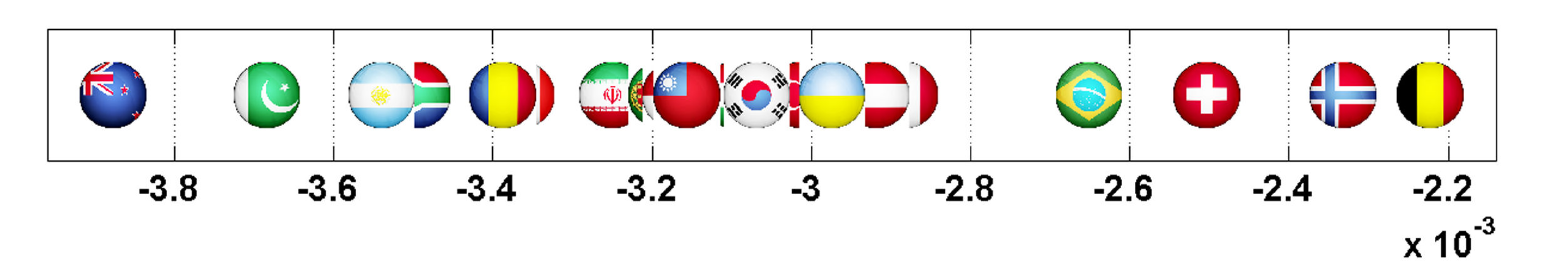}
\caption{{\bf Axial representation of $\bar{D}$ for link modification from GB to FR.}
(not shown $\bar{D}(GB)=-0.00403$ and $\bar{D}(FR)=0.0368$).}
\label{Fig22}
\end{figure}

\begin{figure}[!ht]
\centering
\includegraphics[scale=0.7]{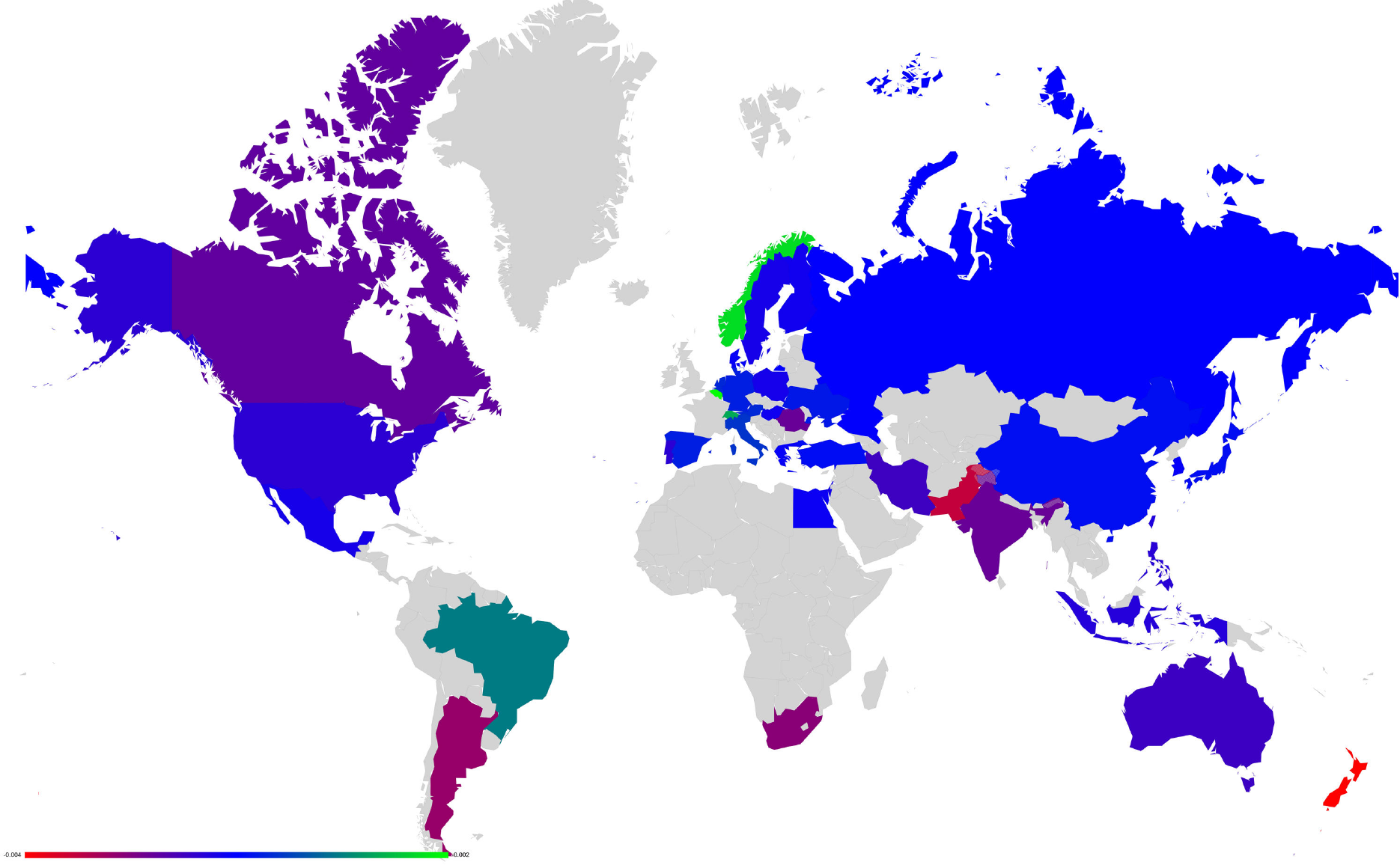}
\caption{{\bf Map representation of $\bar{D}$ for link modification from GB to FR.}
For linked countries  $\bar{D}$ is not shown.
}
\label{Fig23}
\end{figure}

\paragraph*{United Kingdom to France.} The modification of this link gives the most strong effect on
New Zealand (see Figs.~\ref{Fig22},~\ref{Fig23}). Indeed,
referring to New Zealand Ministry of Foreign Affairs and Trade~\cite{uknz}, UK is the top destination for New Zealand's goods and services exports within the EU, and a base for New Zealand companies doing business in Europe. According to the statistics of March 2015, the total trade in goods between the two countries is \$2,807 billion. New Zealand works closely with UK to face terrorism: strategic dialogue talks on security policy issues with UK are held every year. Also, New Zealand shares important cultural and historical links with UK. For New Zealand, UK is the key to Europe. This means intuitively that New Zealand will be strongly affected by the Brexit. These facts are totally in line with our sensitivity analysis conclusions plotted in Fig~\ref{Fig23} and~\ref{Fig22}. In order to face the consequences of Brexit together, UK and NZ have started a serious discussion as mentioned in ~\cite{independentbrexit,hubbrexit}.

\FloatBarrier

\begin{figure}[!ht]
 \centering
 \includegraphics{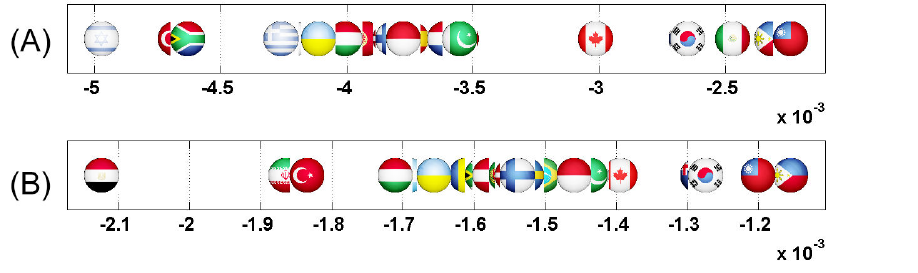}
 \caption{{\bf Axial representation of $\bar{D}$ for links modifications from \{IL and EG\} to US.}
 (A):EG to US (not shown $\bar{D}(EG)=-0.0080$ and $\bar{D}(US)=0.0252$). 
(B):IL to US (not shown $\bar{D}(IL)=-0.0041$ and $\bar{D}(US)=0.0108$).}
 \label{Fig24}
 \end{figure}

\begin{figure}[!ht]
 \centering
 \includegraphics[scale=0.7]{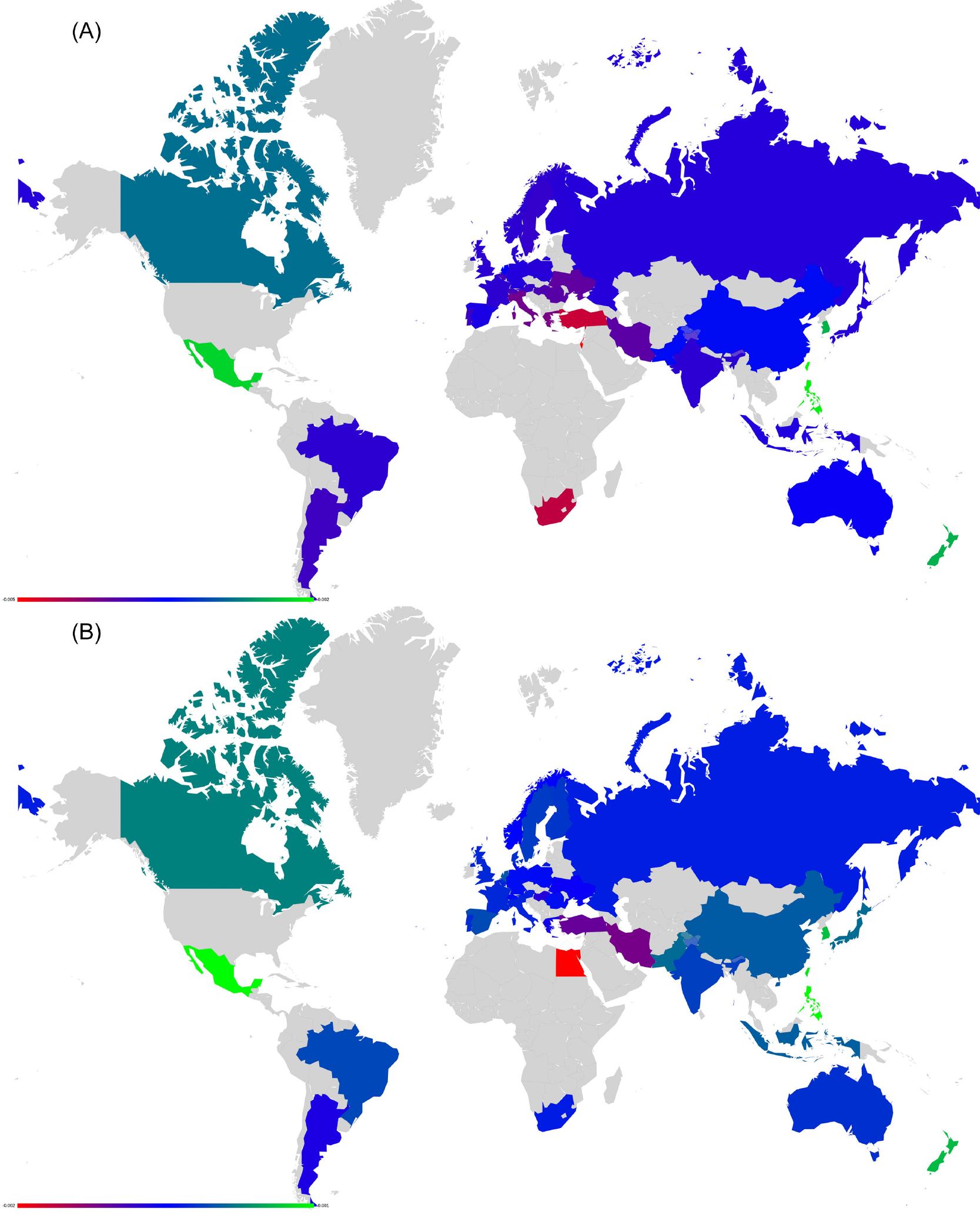}
 \caption{{\bf Map representation of $\bar{D}$ for links modifications from \{IL and EG\} to US.}
 (A):EG to US. (B):IL to US. For linked countries  $\bar{D}$ is not shown.}
 \label{Fig25}
 \end{figure}

\paragraph{US-Israel-Egypt.} 
The Arab-Israeli relationship has been conflicting ever since the Jewish community has shown interest in establishing a nation-state in Palestine. The 1917 Balfour Declaration favored the establishment of a Jewish national home in Palestine and US supported it ~\cite{balfour}. On November 29 1947, the United Nations General Assembly adopted the partition resolution number 181~\cite{181} that would divide Palestinian territory into Jewish and Arab states. Again, US stood aside Israel in supporting the United Nations resolution. Palestinians (and Arabs in general) denounced the partition. Since then, Arab-Israeli did combat in five major wars (1948, 1956, 1967, 1973 and 1982) with Egypt the leader of Arab side in 3 out of 5 wars. Even though the Camp David Accords ~\cite{campdavid} between Egypt and Israel were signed on September 17, 1978  followed by a peace treaty on March 26, 1979 ~\cite{peace} (both being signed in US and witnessed by Jimmy Carter), the relationship is still conflicting. It has been called the ``cold peace''. On the other side, Israeli-US relations are getting stronger according to Jeremy M. Sharp ~\cite{USidtoIsrael}: Israel is the largest cumulative recipient of US foreign aid since World War II. Our results show in Fig~\ref{Fig24} and~\ref{Fig25} that Egypt and Israel will be the most affected countries if the other one gets closer to US.

\FloatBarrier

\begin{figure}[!ht]
\centering
\includegraphics{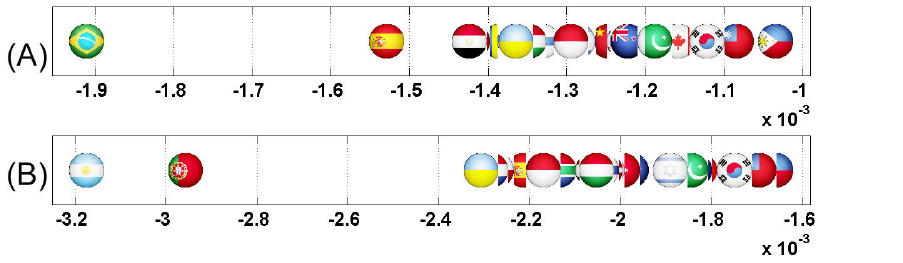}
\caption{{\bf Axial representation of $\bar{D}$ for link modifications from \{AR and BR\} to US.}
(A): AR to US (not shown $\bar{D}(AR)=-0.0050$ and $\bar{D}(US)=0.0094$). 
(B): BR to US (not shown $\bar{D}(BR)=-0.0074$ and $\bar{D}(US)=0.0149$).}
\label{Fig26}
\end{figure}
\begin{figure}[!ht]
\centering
\includegraphics[scale=0.7]{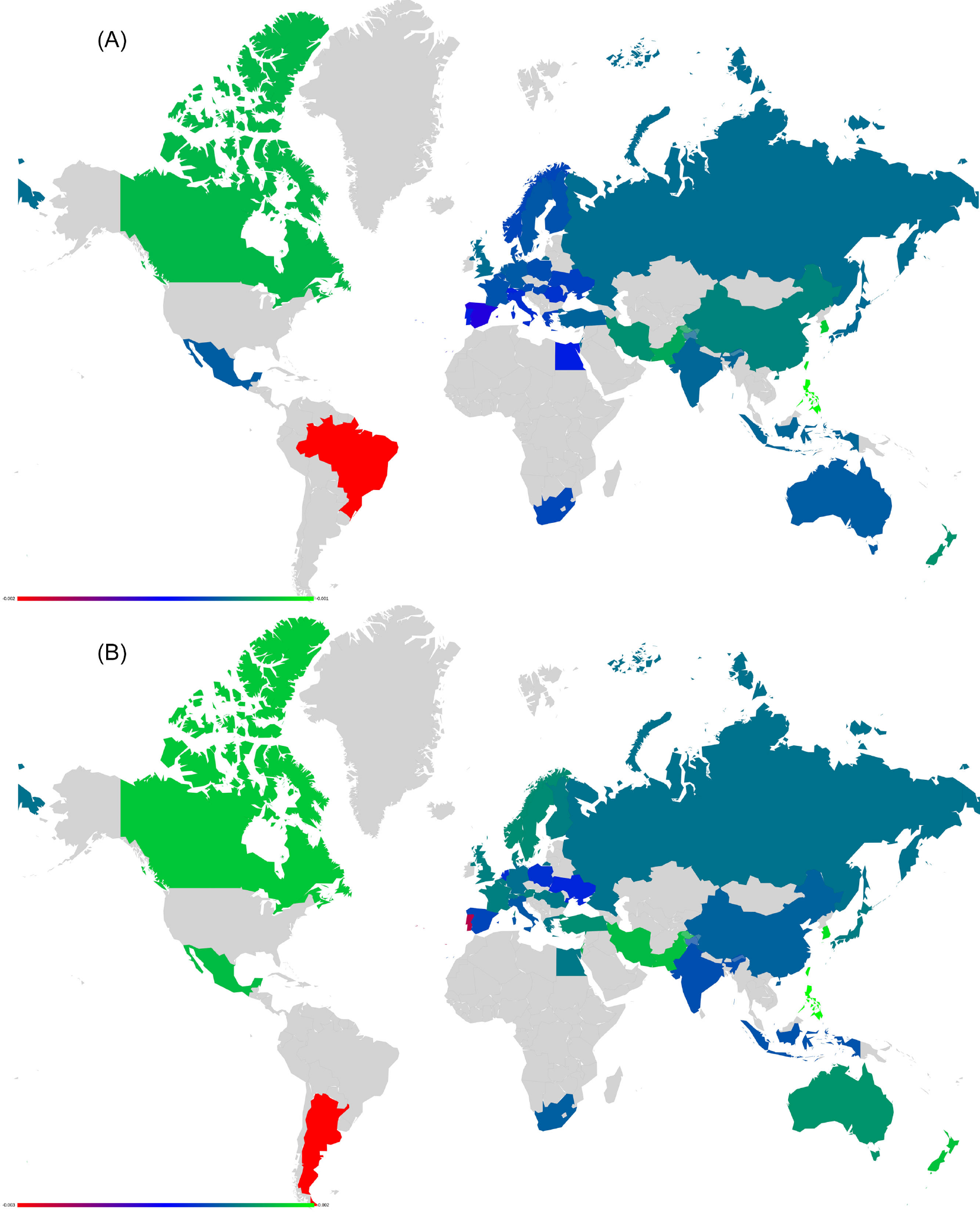}
\caption{{\bf Map representation of $\bar{D}$ for link modifications from \{AR and BR\} to US.}
(A): AR to US. (B): BR to US. For linked countries  $\bar{D}$ is not shown. }
\label{Fig27}
\end{figure}

\paragraph*{Argentina and Brazil} Their relationship~\cite{ArgentinaBrazil} includes all possible fields: economy, history, culture, trade and social structure. As members of the Mercosur sub-regional bloc, Argentina and Brazil relationship offers free trade and fluid movement of goods, people, and currency. Besides that, a Nuclear Cooperation between these two countries was signed on July 18, 1991 and the Brazilian-Argentine Agency for Accounting and Control of Nuclear Materials (ABACC) was created as a binational safeguard organization. Comparing our results (shown in Fig~\ref{Fig26} and~\ref{Fig27}) with these facts of strong relationship between Argentina and Brazil, we find that any unilateral rapprochement between Argentina or Brazil to US will negatively affect the other country. 

\FloatBarrier

\FloatBarrier

%
%
%
%
%

\subsubsection*{Relationship imbalance analysis}

\begin{figure}[!ht]
\centering
\includegraphics[scale=0.9]{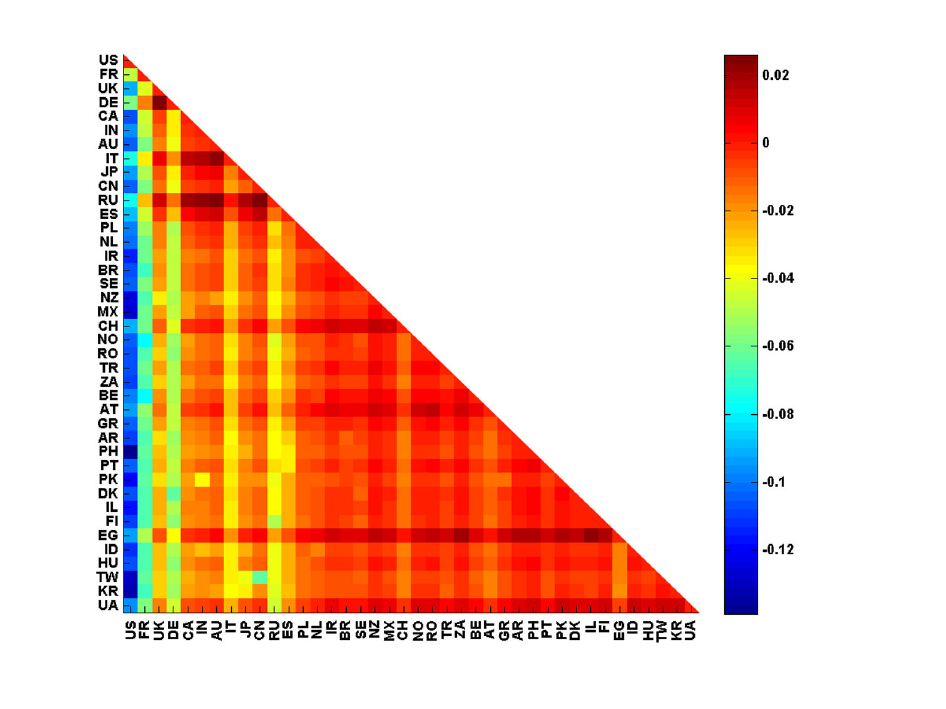}
\caption{{\bf Relationship imbalance analysis: F-representation for 27 EU network.} 
$F(a,b)$ is given by the colorbar. X-axis and Y-axis represent $a$ and $b$ respectively. If $F(a,b)$ is negative, nation $a$ has more influence on nation $b$ than $b$ on $a$.  
}
\label{Fig28}
\end{figure}

Relationship imbalance analysis has been derived for all pairs of 40 countries following Eq~\eqref{eq_F} as well.
Fig~\ref{Fig28} shows a density plot of $F(a,b)$. 
US is clearly the dominant country among all other 39 countries chosen worldwide. Also, Fig~\ref{Fig28} shows that some countries have a strong influences such as France, Germany, Russia, China and Egypt. Germany and France are the two main players of European Union. Russia has an long history of sovereignty over eastern Europe and Northern Asia, economically, politically and culturally. Egypt plays a central role in the middle east. China, with its large population and strong economy, is dominating several countries. However, its role may be underestimated since no Chinese Wikipedia edition is accounted for in our study. 

\FloatBarrier
\section*{Discussion}

This work offers a new perspective for future geopolitics studies. It is possible to extract from multi-cultural Wikipedia networks a global understanding of the interactions between countries at a global, continental or regional scale. Reduced Google matrix theory has been shown to capture hidden interactions among countries, resulting in new knowledge on geopolitics. Results show that our sensitivity analysis captures the importance of relationships on network structure. This analysis relies on the reduced Google matrix and leverages its capability of concentrating all Wikipedia knowledge in a small stochastic matrix. We stress that the obtained sensitivity of geopolitical relations between two countries and its influence on other world countries is obtained on a pure mathematical statistical analysis without any direct appeal to political, economical and social  sciences. 

\section*{Acknowledgments}
This work was supported by APR 2015 call of University of Toulouse and by R\'egion Occitanie (project GOMOBILE), MASTODONS-2016 CNRS project APLIGOOGLE, and EU CHIST-ERA MACACO project ANR-13-CHR2-0002-06.

\nolinenumbers

%
%
%
\FloatBarrier

\end{document}